\documentclass[smallextended]{amsart}
\usepackage{etoolbox}
\usepackage{geometry}
\usepackage{graphics}
\newtheorem{theorem}{Theorem}[section]
\newtheorem{lemma}[theorem]{Lemma}
\newtheorem{pro}[theorem]{Proposition}
\newtheorem{cor}[theorem]{Corollary}
\newtheorem{remark}[theorem]{Remark}

\theoremstyle{definition}

\theoremstyle{remark}

\numberwithin{equation}{section}

\begin{document}
\pagestyle{plain}
%\pagenumbering{arabic}
%\bibliographystyle{plain}

\author{Carlos F. Lardizabal and Rafael R. Souza}

\thanks{R.R. Souza is partially supported by FAPERGS (proc. 002063-2551/13-0).}

%Author XXX\textsuperscript{a}\thanks{\noindent\textsuperscript{a}Address of the author XXX \\
%Email...}

\address{Instituto de Matem\'atica - Universidade Federal do Rio Grande do Sul - UFRGS - Av. Bento Gon\c calves 9500 - CEP 91509-900 Porto Alegre, RS, Brazil}
\email{cfelipe@mat.ufrgs.br, rafars@mat.ufrgs.br}

\def\laa{\langle}
\def\raa{\rangle}
\def\qed{\begin{flushright} $\square$ \end{flushright}}
\def\qee{\begin{flushright} $\Diamond$ \end{flushright}}
\def\ov{\overline}

\begin{abstract}
We study a particular class of trace-preserving completely positive maps, called PQ-channels, for which classical and quantum evolutions are isolated in a certain sense. By combining open quantum random walks with a notion of recurrence, we are able to describe criteria for recurrence of the walk related to this class of channels. Positive recurrence for open walks is also discussed in this context.
\end{abstract}

\date{\today}

\makeatletter
\patchcmd{\@maketitle}
  {\ifx\@empty\@dedicatory}
  {\ifx\@empty\@date \else {\vskip3ex \centering\footnotesize\@date\par\vskip1ex}\fi
   \ifx\@empty\@dedicatory}
  {}{}
\patchcmd{\@adminfootnotes}
  {\ifx\@empty\@date\else \@footnotetext{\@setdate}\fi}
  {}{}{}
\makeatother

\title{On a class of quantum channels, open random walks and recurrence}

\maketitle


\section{Introduction}

Completely positive maps (CP) are the basic mathematical objects used to describe open quantum systems. Informally, such maps present both the quantum character of the system together with possible dissipative effects caused by the environment. They also appear in research areas connected to operator algebras \cite{paulsen,takesaki}, dynamical systems \cite{alicki,benatti}, quantum information theory \cite{nielsen,petz} and if a CP map is moreover trace-preserving (CPT), we call it a quantum channel. In this work we are interested in a particular class of channels for which it is possible to obtain a decomposition in terms of a classical, real stochastic matrix $P=(p_{ij})$ and a non-classical evolution, described by a matrix $Q=(q_{ij})$, thus producing the matrix representation, in the 1-qubit case,
\begin{equation}
[\Phi]=\begin{bmatrix} p_{11} & 0 & 0 & p_{12} \\ 0 & q_{11} & q_{12} & 0 \\ 0 & \ov{q_{12}} & \ov{q_{11}} & 0 \\ p_{21} & 0 & 0 & p_{22}\end{bmatrix}
\end{equation}
Not every channel can be written in this form, and the CPT maps which admit such matrix representation will be called {\bf PQ-channels} (to be defined precisely later). $PQ$-channels are computationally simpler than general quantum systems because the main diagonal of the density matrix is not directly affected by its quantum portion (coherences), and conversely: let $\rho$ be a density matrix and write the iteration $\Phi(\rho)$ in matrix form,
\begin{equation}
[\Phi]vec(\rho)=\begin{bmatrix} p_{11} & 0 & 0 & p_{12} \\ 0 & q_{11} & q_{12} & 0 \\ 0 & \ov{q_{12}} & \ov{q_{11}} & 0 \\ p_{21} & 0 & 0 & p_{22}\end{bmatrix}\begin{bmatrix} \rho_{11} \\ \rho_{12} \\ \ov{\rho_{12}} \\ \rho_{22}\end{bmatrix}=\begin{bmatrix} \rho_{11}p_{11}+\rho_{22}p_{22} \\ q_{11}\rho_{12}+q_{12}\ov{\rho_{12}} \\ \ov{q_{12}}\rho_{12}+\ov{q_{11}}\ov{\rho_{12}} \\ \rho_{11}p_{21}+\rho_{22}p_{22}\end{bmatrix},\;\;\;\rho=\begin{bmatrix} \rho_{11} & \rho_{12} \\ \ov{\rho_{12}} & \rho_{22} \end{bmatrix}
\end{equation}
Above $vec(\rho)$ means $\rho$ written in vector form. We immediately realize that the diagonal entries of such iteration (top and bottom entries of the vector form) depend only on $\rho$ itself and the $p_{ij}$, whereas the antidiagonal part (second and third entries) depend only on $\rho$ and the $q_{ij}$.

\medskip

This work has two main objectives. First, we show that the class of PQ-channels is rich enough to provide nonclassical phenomena and at the same time requires an analysis which is often simpler than the ones imposed by general open (or closed) interactions. In order to encourage further research on this matter, we study some probabilistic aspects of CP maps for which a description in this class of channels may shed light on the structure of more general cases. The second objective is to combine a model of open quantum random walk \cite{attal}, described in terms of CP maps, with a notion of recurrence \cite{werner} and study some of its properties in our context. In this setting we are able to prove a recurrence criteria for  PQ-channels. We are also able to discuss a notion of positive recurrence for open quantum walks. A basic fact which allows a consistent analysis of matrix representations is that such structure is independent of the Kraus matrices chosen for a given channel. We will restrict ourselves to the finite dimensional case, meaning that the CP maps considered act on the state space of matrices $M_d=M_d(\mathbb{C})$. Emphasis on 1-qubit PQ-channels will be given but results on larger spaces are described, and general statements are established.

\medskip

This work is structured as follows. In section \ref{sec2} we review basic facts on completely positive maps and matrix representations. In section \ref{secpq} we define PQ-channels and give some basic characterizations. In section \ref{oqrw} we describe open quantum random walks induced by PQ-channels, following the setting presented in \cite{attal}. Inspired by the definition of recurrence given in \cite{werner}, we define recurrence for open walks, and establish criteria for recurrence when the transitions are induced by certain 1-qubit PQ-channels. In section \ref{section_station} we study positive recurrence for open quantum walks. We note that the structure of some proofs is inspired by classical results. In section \ref{openq} we state open questions and for reference we include in the appendix examples of matrix representations of PQ-channels.

\section{Quantum channels and matrix representations}\label{sec2}

We make a brief review on completely positive maps \cite{nielsen,petz,watrous}. For simplicity we will restrict ourselves to the space $M_d=M_d(\mathbb{C})$ of order $d$ matrices. A hermitian matrix $A:\mathbb{C}^d\to\mathbb{C}^d$ is positive, denoted by $A\geq 0$, if $\langle Av,v\rangle\geq 0$, for all $v\in\mathbb{C}^d$. We say $\rho\in M_d(\mathbb{C})$ is a density matrix if $\rho\geq 0$ and $tr(\rho)=1$. Let $\Phi: M_d(\mathbb{C})\to M_d(\mathbb{C})$ be linear. We say $\Phi$ is a {\bf positive} operator whenever $A\geq 0$ implies $\Phi(A)\geq 0$. In quantum mechanics, we are mostly interested in $\Phi$ that are not only positive over $M_d$ (system A), but also that $I\otimes \Phi$ is also positive, where $I$ is the identity operator on any other choice of component (system B). More precisely, define for each $k\geq 1$, $\Phi_k: M_k(M_n(\mathbb{C}))\to M_k(M_n(\mathbb{C}))$, $\Phi_k(A)=[\Phi(A_{ij})],\;\;\; A\in M_k(M_n(\mathbb{C})), \; A_{ij}\in M_n(\mathbb{C})$. We say $\Phi$ is $k$-positive if $\Phi_k$ is positive, and we say it is {\bf completely positive} (CP) if $\Phi_k$ is positive for every $k=1, 2,\dots$. It is well-known that a map $\Phi$ is CP if and only if it can be written in the Kraus form
\begin{equation}\label{krausform}
\Phi(\rho)=\sum_i V_i\rho V_i^*,
\end{equation}
where the $V_i$ are linear operators (Kraus matrices) \cite{petz}. We say $\Phi$ is {\bf trace-preserving} if $tr(\Phi(\rho))=tr(\rho)$ for all $\rho\in M_d(\mathbb{C})$, which is equivalent to $\sum_i V_i^*V_i=I$. We say $\Phi$ is {\bf unital} (or {\bf doubly stochastic}) if $\Phi(I)=I$, which is equivalent to $\sum_i V_iV_i^*=I$. In this work we are interested in completely positive trace-preserving (CPT) maps (also called {\bf channels}) acting on a finite dimensional space. Also recall that if $A\in M_d(\mathbb{C})$ there is the corresponding vector representation $vec(A)$ associated to it, given by stacking together the matrix rows. For instance, if $d=2$,
\begin{equation}
A=\begin{bmatrix} a_{11} & a_{12} \\ a_{21} & a_{22}\end{bmatrix}\;\;\;\Rightarrow \;\;\; vec(A)=\begin{bmatrix} a_{11} & a_{12} & a_{21} & a_{22}\end{bmatrix}^T.
\end{equation}
The $vec$ mapping satisfies $vec(AXB^T)=(A\otimes B)vec(X)$ for any $A, B, X$ square matrices \cite{hornjohnson} so in particular, $vec(VXV^*)=vec(VX\ov{V}^T)=(V\otimes \ov{V})vec(X)$,
from which we can obtain the {\bf matrix representation} $[\Phi]$ for the CP map (\ref{krausform}):
\begin{equation}\label{matrep}
[\Phi]=\sum_{i} V_{i}\otimes \ov{V_{i}}
\end{equation}

\begin{remark}\label{bas_remark}
We recall the well-known fact that the matrix representation of a CPT map $\Phi: M_d(\mathbb{C})\to M_d(\mathbb{C})$ is independent of the Kraus representation considered. This is a consequence of the unitary equivalence of Kraus matrices for a given quantum channel \cite{petz} and is necessary for a consistent analysis in the following sections.
\end{remark}

\section{PQ-channels}\label{secpq}

{\bf Definition.} Let $\Phi: M_2(\mathbb{C})\to M_2(\mathbb{C})$ be a CPT map. If the associated matrix representation $[\Phi]$ is given by
\begin{equation}\label{def_pq}
[\Phi]=\begin{bmatrix} p_{11} & 0 & 0 & p_{12} \\ 0 & q_{11} & q_{12} & 0 \\ 0 & \ov{q_{12}} & \ov{q_{11}} & 0 \\ p_{21} & 0 & 0 & p_{22} \end{bmatrix},
\end{equation}
where  $P=(p_{ij})$ is an order 2 stochastic matrix, $q_{ij}\in\mathbb{C}$, we say that $\Phi$ is an {\bf order 2 PQ-channel}. The particular aspect of the $q_{ij}$ entries (e.g. the terms $q_{11}$ and $\ov{q_{11}}$ in the main diagonal) are due to the multiplication rule given by (\ref{matrep}).
For instance, let $\Lambda(X)=V_1XV_1^*+V_2XV_2^*$ be trace-preserving and write
\begin{equation}
V_1=\begin{bmatrix} x_{11} & x_{12} \\ x_{21} & x_{22} \end{bmatrix},\;\;\; V_2=\begin{bmatrix} y_{11} & y_{12} \\ y_{21} & y_{22} \end{bmatrix}.
\end{equation}
Then we obtain
\begin{equation}
[\Lambda]=\begin{bmatrix} |x_{11}|^2+|y_{11}|^2 & x_{11}\ov{x_{12}}+y_{11}\ov{y_{12}} & \ov{x_{11}}x_{12}+\ov{y_{11}}y_{12} & |x_{12}|^2+|y_{12}|^2 \\ x_{11}\ov{x_{21}}+y_{11}\ov{y_{21}} & x_{11}\ov{x_{22}}+y_{11}\ov{y_{22}} & x_{12}\ov{x_{21}}+y_{12}\ov{y_{21}} & x_{12}\ov{x_{22}}+y_{12}\ov{y_{22}} \\ \ov{x_{11}}x_{21}+\ov{y_{11}}y_{21} & \ov{x_{12}}x_{21}+\ov{y_{12}}y_{21} & \ov{x_{11}}x_{22}+\ov{y_{11}}y_{22} & \ov{x_{12}}x_{22}+\ov{y_{12}}y_{22} \\ |x_{21}|^2+|y_{21}|^2 & x_{21}\ov{x_{22}}+y_{21}\ov{y_{22}} & \ov{x_{21}}x_{22}+\ov{y_{21}}y_{22} & |x_{22}|^2+|y_{22}|^2
\end{bmatrix},
\end{equation}
which is of the form
\begin{equation}\label{letterform}
[\Lambda]=\begin{bmatrix} |a| & b & \ov{b} & |c| \\ d & e & f & g \\
\ov{d} & \ov{f} & \ov{e} & \ov{g} \\ |h| & j & \ov{j} & |k|
\end{bmatrix},
\end{equation}
where the letters indicate complex numbers, the bars $|\cdot|$ emphasizing that such entry is a nonnegative real number. More generally, if $\Phi: M_d(\mathbb{C})\to M_d(\mathbb{C})$ is a CPT map with associated order $d^2$ matrix representation $[\Phi]$ given by
\begin{equation}\label{largegeneraldef}
\begin{bmatrix}
p_{11} & \cdots  &  p_{12} & \cdots  & p_{13} & \cdots & \cdots & p_{1(d-1)} & \cdots & p_{1d} \\
\vdots & Q_{11} & \vdots & Q_{12}  & \vdots & \cdots & \cdots & \vdots & Q_{1(d-1)} & \vdots\\
p_{21} & \cdots  &  p_{22} & \cdots  & p_{23} & \cdots & \cdots & p_{2(d-1)} & \cdots & p_{2d} \\
\vdots &  & \vdots &   & \vdots & \cdots & \cdots & \vdots & & \vdots\\
\vdots &  & \vdots &   & \vdots & \cdots & \cdots & \vdots & & \vdots\\
p_{(d-1)1} & \cdots  &  p_{(d-1)2} & \cdots  & p_{(d-1)3} & \cdots & \cdots & p_{(d-1)(d-1)} & \cdots & p_{(d-1)d} \\
\vdots & Q_{(d-1)1} & \vdots & Q_{(d-1)2}  & \vdots & \cdots & \cdots & \vdots & Q_{(d-1)(d-1)} & \vdots\\
p_{d1} & \cdots  &  p_{d2} & \cdots  & p_{d3} & \cdots & \cdots & p_{d(d-1)} & \cdots & p_{dd} \\
\end{bmatrix}
\end{equation}
where the $Q_{ij}=(q_{ij})$ are order $d$ matrices surrounded by null entries (represented by dots around the $Q_{ij}$), and $P=(p_{ij})$ is an order $d$ stochastic matrix, then we say that $\Phi$ is an {\bf order d PQ-channel}. It is easy to see that the composition (matrix product) of PQ-channels is also of this kind and if $\Phi(X)=\sum_i V_i XV_i^*$  is a PQ-channel, then so is $\Phi^*(X)=\sum_i V_i^*XV_i$. If $Q_{ij}=0$, for all $i,j=1, 2, \dots d-1$ then we say that $\Phi$ is a {\bf Markov channel}. We say that the entries $p_{ij}$ correspond to the {\bf classical part} of the channel and the $q_{ij}$ form the {\bf quantum part}. Also, we have that $|q_{ij}|\leq 1$ and, in general, the $Q_{ij}$ are complex matrices,  which we call {\bf $Q$-blocks}. A routine calculation shows that the bit flip, phase-flip, bit-phase flip, amplitude damping, depolarizing channel, and phase damping channels \cite{nielsen} are 1-qubit PQ-channels (see the appendix). As another example consider the general unitary matrix in $M_2$,
\begin{equation}\label{unitarypq}
U=\begin{bmatrix} e^{i\alpha}cos(\theta) & e^{i\gamma}sin(\theta) \\ -e^{i(\beta-\gamma)}sin(\theta) & e^{i(\beta-\alpha)}cos(\theta)
\end{bmatrix},\;\;\; \alpha,\beta,\gamma,\theta\in\mathbb{R}
\end{equation}
Then if $\Phi_U(X)=UXU^*$ then $[\Phi_U]$ will represent a PQ-channel if and only if $\theta=k\pi$ or $\theta=k\pi/2$, $k\in\mathbb{Z}$ (see Proposition \ref{pprroo1}). This example represents the fact that classical and quantum variables increase their interactions (under an algebraic point of view) whenever a generic unitary action is performed in the system, but if the action performed is a PQ-channel, then the behavior of such correlations is simpler. This is not a new idea, instead it is an elementary consequence of the algebraic structure we are interested here.

\begin{remark}
In this work we will suppose for simplicity that the channels act on a fixed matrix space $M_n(\mathbb{C})$ and that $[\Phi]$ is written in the canonical basis both in the domain and range of the map. On the other hand, if a change of basis is performed, the PQ property might be lost (or gained). For instance, unitary matrices in general do not induce PQ maps when written in the canonical basis (see eq. (\ref{unitarypq})), but these are diagonal when written in terms of its associated basis of eigenvectors. See also Remark \ref{bas_remark}.
\end{remark}

\begin{pro}\label{pprroo1}
A 1-qubit CPT map $\Phi$ is a PQ-channel if and only if $\Phi$ admits a Kraus representation consisting exclusively of matrices which are diagonal or anti-diagonal, both kinds possibly appearing.
\end{pro}
{\bf Proof.} If all order 2 Kraus matrices are diagonal or antidiagonal, then it is a simple matter to check the associated matrix representation and see that it has the form of a PQ-channel. Conversely, let $\Phi$ be a PQ-channel, so we are given some Kraus decomposition $\Phi(\cdot)=\sum_i V_i\cdot V_i^*$. For simplicity of notation, assume that we have 2 Kraus matrices. The general case of $k$ Kraus matrices follows by an identical reasoning on each matrix. Write
\begin{equation}
V_1=\begin{bmatrix} x_{11} & x_{12} \\ x_{21} & x_{22}\end{bmatrix},\;\;\; V_2=\begin{bmatrix} y_{11} & y_{12} \\ y_{21} & y_{22}\end{bmatrix}
\end{equation}
Then, as $\Phi$ is by assumption a $PQ$-channel, we must have
\begin{equation}
[\Phi]=\begin{bmatrix} |x_{11}|^2+|y_{11}|^2 & 0 & 0 & |x_{12}|^2+|y_{12}|^2 \\ 0 & x_{11}\ov{x_{22}}+y_{11}\ov{y_{22}} & x_{12}\ov{x_{21}}+y_{12}\ov{y_{21}} & 0 \\ 0 & \ov{x_{12}}x_{21}+\ov{y_{12}}y_{21} & \ov{x_{11}}x_{22}+\ov{y_{11}}y_{22} & 0 \\ |x_{21}|^2+|y_{21}|^2 & 0 & 0 & |x_{22}|^2+|y_{22}|^2
\end{bmatrix},
\end{equation}
(also recall Remark \ref{bas_remark} on the independence of the matrix representation with respect to the choice of Kraus matrices). Let
\begin{equation}
B_1=\begin{bmatrix} x_{11} & 0 \\ 0 & x_{22}\end{bmatrix},\;\;\; B_2=\begin{bmatrix} 0 & x_{12} \\ x_{21} & 0\end{bmatrix},\;\;\;C_1=\begin{bmatrix} y_{11} & 0 \\ 0 & y_{22}\end{bmatrix},\;\;\; C_2=\begin{bmatrix} 0 & y_{12} \\ y_{21} & 0\end{bmatrix}
\end{equation}
and note that
\begin{equation}
[B_1]+[B_2]+[C_1]+[C_2]=\begin{bmatrix} |x_{11}|^2+|y_{11}|^2 & 0 & 0 & |x_{12}|^2+|y_{12}|^2 \\ 0 & x_{11}\ov{x_{22}}+y_{11}\ov{y_{22}} & x_{12}\ov{x_{21}}+y_{12}\ov{y_{21}} & 0 \\ 0 & \ov{x_{12}}x_{21}+\ov{y_{12}}y_{21} & \ov{x_{11}}x_{22}+\ov{y_{11}}y_{22} & 0 \\ |x_{21}|^2+|y_{21}|^2 & 0 & 0 & |x_{22}|^2+|y_{22}|^2
\end{bmatrix}=[\Phi]
\end{equation}

\qed

If $d>2$ then there are PQ-channels which are not spanned by diagonal and antidiagonal matrices alone. In order to generate all possible channels of such kind, we consider instead the set
\begin{equation}\label{pq_deff}
PQ_d=\{ A=(a_{ij})\in M_d(\mathbb{C}): a_{kl}\neq 0 \Rightarrow a_{xl}=0, a_{ky}=0, \forall x\neq k, y\neq l\},
\end{equation}
that is, the set of matrices such that if a given entry is nonzero, then it is the only nonzero element in its row and column. Or still, the set of matrices which are a permutation of some diagonal matrix. We call matrices in $PQ_d$ a {\bf PQ-matrix}. If $d=2$ then this set equals the set of diagonal and antidiagonal matrices.
For $d>2$, it is a simple matter to show that a channel over $M_d(\mathbb{C})$ which admits a Kraus representation consisting exclusively of matrices belonging to $PQ_d$ is a PQ-channel. We believe that the converse is true, but a proof of this statement is an open question. In principle, the problem seems more difficult than the case $d=2$ because of all the possibilities of permutations allowed. The following example illustrates a particular solution.

\medskip

{\bf Example.} Let
\begin{equation}
V_1=\frac{1}{\sqrt{2}}\begin{bmatrix} 0 & -i & 0 \\ i & 0 & 0 \\ 0 & 0 & 0 \end{bmatrix},\;\;\;V_2=\frac{1}{2}\begin{bmatrix} 0 & 0 & -i \\ 0 & 0 & -i \\ i & i & 0 \end{bmatrix},\;\;\;V_3=\frac{1}{2}\begin{bmatrix} 0 & 0 & i \\ 0 & 0 & -i \\ -i & i & 0\end{bmatrix}
\end{equation}
Then a simple calculation shows that $\Phi_{LS}(\cdot):=\sum_i V_i\cdot V_i^*$ is a PQ-channel:
\begin{equation}\label{landaumat}
[\Phi_{LS}]=\begin{bmatrix}
{\bf 0} &  &  &  & {\bf \frac{1}{2}} &  &  &  & {\bf \frac{1}{2}} \\    %1
 & 0 & 0 & -\frac{1}{2} &  & 0 & 0 & 0 &  \\    %2
 & 0 & 0 & 0 &  & 0 & -\frac{1}{2} & 0 &  \\    %3
 & -\frac{1}{2} & 0 & 0 &  & 0 & 0 & 0 &  \\    %4
\bf{\frac{1}{2}}  &  &  &  & \bf{0}  &  &  &  & \bf{\frac{1}{2}} \\    %5
 & 0 & 0 & 0 &  & 0 & 0 & -\frac{1}{2}  &  \\    %6
 & 0 & -\frac{1}{2} & 0 &  & 0 & 0 & 0 &  \\    %7
 & 0 & 0 & 0 &  & -\frac{1}{2} & 0 & 0 &  \\    %8
\bf{\frac{1}{2}}  &  &  &  & \bf{\frac{1}{2}}  &  &  &  & \bf{0} \\    %9
\end{bmatrix}
\end{equation}
The numbers in bold refer to the classical part (matrix P) and the 4 order 3 blocks refer to the quantum part. The blank entries for $[\Phi_{LS}]$ are zeroes, which have been omitted for an easier visualization. This channel is the Landau-Streater channel, described in \cite{landau}. In this example we are able to obtain a Kraus decomposition for $\Phi_{LS}$ in terms of PQ-matrices. In fact, note that $V_1$ is already of the desired form, so we proceed to $V_2$ and $V_3$. One natural attempt for a decomposition of such matrices is to set
\begin{equation}
V_2=V_2^1+V_2^2=\frac{1}{2}\begin{bmatrix} 0 & 0 & -i \\ 0 & 0 & 0 \\ i & 0 & 0 \end{bmatrix}+\frac{1}{2}\begin{bmatrix} 0 & 0 & 0 \\ 0 & 0 & -i \\ 0 & i & 0 \end{bmatrix}
\end{equation}
and
\begin{equation}
V_3=V_3^1+V_3^2=\frac{1}{2}\begin{bmatrix} 0 & 0 & i \\ 0 & 0 & 0 \\ -i & 0 & 0 \end{bmatrix}+\frac{1}{2}\begin{bmatrix} 0 & 0 & 0 \\ 0 & 0 & -i \\ 0 & i & 0 \end{bmatrix}
\end{equation}
from which we get
\begin{equation}
[V_2^1]+[V_2^2]=[V_3^1]+[V_3^2]=\begin{bmatrix}
{\bf 0} &  &  &  & 0 &  &  &  & {\bf \frac{1}{4}} \\    %1
 & 0 & 0 & 0 &  & 0 & 0 & 0 &  \\    %2
 & 0 & 0 & 0 &  & 0 & -\frac{1}{4} & 0 &  \\    %3
 & 0 & 0 & 0 &  & 0 & 0 & 0 &  \\    %4
\bf{0}  &  &  &  & \bf{0}  &  &  &  & \bf{\frac{1}{4}} \\    %5
 & 0 & 0 & 0 &  & 0 & 0 & -\frac{1}{4}  &  \\    %6
 & 0 & -\frac{1}{4} & 0 &  & 0 & 0 & 0 &  \\    %7
 & 0 & 0 & 0 &  & -\frac{1}{4} & 0 & 0 &  \\    %8
\bf{\frac{1}{4}}  &  &  &  & \bf{\frac{1}{4}}  &  &  &  & \bf{0} \\    %9
\end{bmatrix}
\end{equation}
Note the above expression also equals $[V_2]+[V_3]$, where each term in this sum, namely, $[V_2]$ and $[V_3]$, contains nonzero entries in invalid positions for a PQ-channel, but these vanish when they are summed. Then $\Phi_{LS}(\cdot)=V_1\cdot V_1^*+ \sum_{i\in\{2,3\},j\in\{1,2\}} V_i^j\cdot V_i^{j*}$ is a Kraus decomposition in terms of PQ-channels.

\qee

\begin{pro}\label{newprop}
We have:
\begin{enumerate}
\item The set of PQ-channels is convex.
\item The matrix product and the Kronecker product of PQ-matrices is a PQ-matrix.
\item Every CP map majorized by a PQ-channel is induced by PQ-matrices, and is in general non-trace-preserving.
\end{enumerate}
\end{pro}

{\bf Proof.} 1) Immediate. 2) Let $A=P_1D_1$ and $B=P_2D_2$ be two PQ-matrices written as a permutation of a diagonal matrix ($P_i$, $D_i$, $i=1,2$, respectively). Then we can write $AB=P_1D_1P_2D_2=P_1(D_1P_2D_2)=P_1C$, which consists of a permutation of $C$, which in turn is a permutation of a diagonal matrix. To see this, it is enough to notice that $D_1$ is diagonal so that the entries $(i,j)$ for $C$ which are nonzero are the same as for $P_2D_2=B$ (assuming $D_1$ has only nonzero entries in its diagonal; the general case is treated similarly). As for Kronecker product, write for $A$, $B$ PQ-matrices, $A\otimes B=P_1D_1\otimes P_2D_2=(P_1\otimes P_2)(D_1\otimes D_2)$ and use the fact that the Kronecker product of permutations is also a permutation, and analogously for diagonal matrices. 3) This is a consequence of a result due to Raginsky \cite{raginsky}, Theorem III.5 together with Remark \ref{bas_remark}.

\qed

For a 1-qubit PQ-channel,
\begin{equation}
[\Phi]=\begin{bmatrix} p_{11} & 0 & 0 & p_{12} \\ 0 & q_{11} & q_{12} & 0\\ 0 & \ov{q_{12}} & \ov{q_{11}} & 0 \\ p_{21} & 0 & 0 & p_{22}\end{bmatrix}, \;\;\;p_{ij}\geq 0, \;\;\; p_{11}+p_{21}=p_{12}+p_{22}=1,
\end{equation}
we define
$$[P]:=\begin{bmatrix} p_{11} & 0 & 0 & p_{12} \\ 0 & 0 & 0 & 0\\ 0 & 0 & 0 & 0 \\ p_{21} & 0 & 0 & p_{22}\end{bmatrix},\;\;\; [Q]:=\begin{bmatrix} 0 & 0 & 0 & 0 \\ 0 & q_{11} & q_{12} & 0\\ 0 & \ov{q_{12}} & \ov{q_{11}} & 0 \\ 0 & 0 & 0 & 0\end{bmatrix}$$
Then the proof of the following is straightforward:
\begin{pro} %MAPLE: cont_ruellec5.mw
Let $\Phi$ be a unital 1-qubit PQ-channel with classical part $P$ and quantum part $Q$. Then a) $[P][Q]=[Q][P]=0$; b) $[\Phi]^n=[P]^n+[Q]^n$; c) $exp(t[\Phi])=exp(t([P]+[Q]))=exp(t[P])exp(t[Q])$. Moreover, let $L=\Phi-I$ and let $R$ and $S$ be the classical and quantum parts of $L$. Then d) $[R][S]=[S][R]=0$; e) $[L]^n=[R]^n+[S]^n$; f) $exp(t[L])=exp(t([R]+[S]))=exp(t[R])exp(t[S])$.
\end{pro}

In an analogous way we write $[P]$ and $[Q]$ for a 1-qutrit PQ-channel, i.e., the classical and quantum part matrices and we get the analogous results. The proof of a) and b) are straightforward and the proof of c) follows once again from commutativity of $P$ and $Q$. By defining $L=\Phi-I$ we get the same results, so we have that the above Proposition hold for unital 1-qutrit channels. By induction the above results hold for any unital d-dimensional PQ-channel. We also have:

\begin{cor}
If $\Phi$ is a unital d-dimensional $PQ$-channel then for $L=\Phi-I$ we have that $e^{tL}$ is a CP semigroup which is a PQ-channel for every $t\geq 0$.
\end{cor}
{\bf Proof.} Use Proposition \ref{newprop}, item 2. The fact that $L$ is a valid Lindblad generator for a CP semigroup follows from Cirac and Wolf \cite{wolf}, Lemma 1.

\qed

{\bf Example: Unitary decomposition of 1-qubit PQ-channels}. Consider a unital 1-qubit PQ-channel. Then it admits, by Birkhoff's theorem \cite{landau}, a decomposition in terms of unitary matrices. We show explicitly such decomposition with unitaries which are diagonal or antidiagonal. Let $\Phi_D(X)=\frac{1}{2}(U_1XU_1^*+U_2XU_2^*)$, where
\begin{equation}
U_1=\begin{bmatrix} e^{ia} & 0 \\ 0 & e^{ib}\end{bmatrix},\;\;\; U_2=\begin{bmatrix} e^{ic} & 0 \\ 0 & e^{id}\end{bmatrix},\;\;\; a, b, c, d\in\mathbb{R}
\end{equation}
Let $\Phi_A(X)=\frac{1}{2}(U_3XU_3^*+U_4XU_4^*)$, where
\begin{equation}
U_3=\begin{bmatrix} 0& e^{if} \\ e^{ig} & 0 \end{bmatrix},\;\;\; U_4=\begin{bmatrix} 0 & e^{ih} \\ e^{ij} & 0 \end{bmatrix},\;\;\; f, g, h, j\in\mathbb{R}
\end{equation}
Let
\begin{equation}\label{mu5}
[\Phi]=\begin{bmatrix} p_{11} & 0 & 0 & p_{12} \\ 0 & x+iy & z+iw & 0 \\ 0 & z-iw & x-iy & 0 \\ p_{12} & 0 & 0 & p_{11} \end{bmatrix}
\end{equation}
be a unital 1-qubit PQ-channel (recall $p_{12}=p_{21}$ since P is bistochastic in this case). Define
\begin{equation}\label{gen1pqc}
\Phi(X)=\frac{p_{11}}{2}\Phi_D(X)+\frac{p_{12}}{2}\Phi_A(X)
\end{equation}
where $\Phi_D$ and $\Phi_A$ are as above.
Then
\begin{equation}
[\Phi]=\begin{bmatrix} p_{11} & 0 & 0 & p_{12} \\ 0 & \frac{1}{2}p_{11}(e^{i(a-b)}+e^{i(c-d)}) & \frac{1}{2}p_{12}(e^{i(f-g)}+e^{i(h-j)}) & 0 \\ 0 & \frac{1}{2}p_{12}(e^{-i(f-g)}+e^{-i(h-j)}) & \frac{1}{2}p_{11}(e^{-i(a-b)}+e^{-i(c-d)})& 0 \\ p_{12} & 0 & 0 & p_{11}\end{bmatrix}=p_{11}[\Phi_D]+p_{12}[\Phi_A]
\end{equation}
Hence, with a suitable choice of $a, b, c, d, f, g, h, j$, equation (\ref{gen1pqc}) describes any unital 1-qubit PQ-channel as a convex sum of unitary matrices which are diagonal or antidiagonal.

\qee

To conclude this section we state the following Proposition, which is a consequence of the fact that the spectral behavior of PQ-channels is divided into blocks, that is, one may look separately for the eigenvalues of the classical part $P$ and quantum part $Q$. For a given metric $\Vert\cdot \Vert$ (e.g. trace distance), recall that a quantum channel $\Phi$ is {\bf ergodic} if it admits a unique fixed density matrix and we say $\Phi$ is {\bf mixing} if there exists a unique density matrix $\rho_0$ such that $\Vert \Phi^n(\rho)-\rho_0\Vert\to 0$ as $n\to\infty$, for all $\rho$ density \cite{burgarth0,burgarth}. This kind of result should be compared with other known asymptotic results on quantum channels \cite{petulante,novotny1}.

\begin{pro}
a) A PQ-channel with classical and quantum parts $P$ and $Q$ is ergodic if and only if $P$ has a unique fixed point and $Q$ has no fixed points. b) No mixed-permutation channel is ergodic. c) If a 1-qubit unital PQ-channel $\Phi$ has real Q part then $\Phi^*\Phi=\Phi\Phi^*$ and so it is unitarily diagonalizable as a linear operator. d) If a 1-qubit random unitary channel is ergodic then there must be a unitary Kraus matrix which is not diagonal. e) Let $\Phi$ be a 1-qubit unital channel with real $Q$ part. Then $\Phi$ is mixing if and only if $\Phi^*\Phi$ is ergodic.
\end{pro}
{\bf Proof.} a) Note that the spectra of PQ-channels is the union of the spectra of classical and quantum parts $P$ and $Q$. b) Note that the quantum part $Q$ of a such channel is a stochastic matrix (see Example on mixed-permutation channels) and such matrices always have 1 as eigenvalue. c) Note that in this case $P$ and $Q$ are both hermitian. d) The result follows immediately from Burgarth et al. \cite{burgarth0}, Corollary 3, since diagonal matrices always commute. e) This follows from c) and Burgarth et al. \cite{burgarth0}, Theorem 13.

\qed

\section{Open quantum random walks induced by PQ-matrices and recurrence}\label{oqrw}

In this section we study a model of open quantum random walks (OQRW) on $\mathbb{Z}$. Following the setting described in \cite{attal}, let $\mathcal{K}$ denote a separable Hilbert space and let $\{|i\rangle\}_{i\in \mathbb{Z}}$ be an orthonormal basis for such space. Informally, this will describe the space of vertices. Let $\mathcal{H}$ be another Hilbert space, which will describe the degrees of freedom given at each point of $\mathbb{Z}$. Then we will consider the space $\mathcal{H}\otimes\mathcal{K}$. For each edge we associate a bounded operator $B_j^i$ on $\mathcal{H}$. This operator describes the effect of passing from $j$ to $i$. We will assume that for each $j$, $\sum_i B_j^{i*}B_j^{i}=I$, where, if infinite, such series is strongly convergent. This constraint is to be understood as follows: the sum of all the effects leaving the site $j$ is $I$. We will consider density matrices on $\mathcal{H}\otimes\mathcal{K}$ with the particular form $\rho=\sum_i\rho_i\otimes |i\rangle\langle i|$, assuming that $\sum_i tr(\rho_i)=1$.
%That is, we initially consider mixtures of initial states on each site, but no mixing between sites.
 For a given initial state of such form, we define
\begin{equation}
\Omega(\rho):=\sum_i\Big(\sum_j B_j^i\rho_j B_j^{i*}\Big)\otimes |i\rangle\langle i|
\end{equation}
Above we assume that $\sum_i B_j^{i*}B_j^{i}=I$ and note that by defining $M_j^i=B_j^i\otimes |i\rangle\langle j|$ then we can write
\begin{equation}
\Omega(\rho)=\sum_{i,j} M_j^i\rho M_j^{i*}
\end{equation}
and so $\sum_{i,j} M_j^{i*}M_j^{i}=I$, see \cite{attal}. We will say that the quantum channel $\Omega$ constructed in such a way is an {\bf OQRW}. As a consequence of the definition on $M_j^i$ and Proposition \ref{newprop}, item 2, every OQRW induced by PQ-matrices can be seen as an infinite-dimensional version of a PQ-channel. At first we will be interested in random walks for which one can only jump to nearest neighbors. Let $L, R$ two bounded operators on $\mathcal{H}$, associated to a jump to the left or to the right, respectively, and such that
\begin{equation}\label{sum1}
L^*L+R^*R=I,
\end{equation}
and we set for each $i$, $B_i^{i-1}=L$, $B_i^{i+1}=R$, and $B_i^{j}=0$ in all other cases. Let $\rho^{(0)}=\rho_0\otimes |0\rangle\langle 0|$ describe the initial state. After one step we have the state
\begin{equation}
L\rho_0 L^*\otimes |-1\rangle\langle -1|+R\rho_0 R^*\otimes |1\rangle\langle 1|,
\end{equation}
so that the probability of presence in $|-1\rangle$ is $tr(L\rho_0 L^*)$ and the probability of presence in $|1\rangle$ is $tr(R\rho_0 R^*)$. After the second step, the state of the system is
\begin{equation}\label{secondst}
L^2\rho_0L^{2*}\otimes|-2\rangle\langle -2|+(LR\rho_0R^*L^*+RL\rho_0L^*R^*)\otimes |0\rangle\langle 0|+R^2\rho_0 R^{2*}\otimes|2\rangle\langle 2|
\end{equation}
so the associated probabilities for the presence in $|-2\rangle$, $|0\rangle$, $|2\rangle$ are, respectively,
\begin{equation}
tr(L^2\rho_0L^{2*}),\;\;\; tr(LR\rho_0R^*L^*+RL\rho_0L^*R^*),\;\;\; tr(R^2\rho_0 R^{2*})
\end{equation}
This process is repeated indefinitely. Inspired by the recurrence definition for unitary evolutions studied in \cite{werner}, we consider a similar notion for open random walks. Let $\Omega$ denote the open random walk induced, for instance, by matrices $L$ and $R$ satisfying (\ref{sum1}) and let $\rho^{(0)}$ denote the initial state as before (we usually assume that the initial density is positioned at zero). Let $P$ be an operator acting on states of the random walk, in the following way: if $\rho^{(k)}=\Omega^k(\rho^{(0)})$ is a state then $P\rho^{(k)}$ is equal to $\rho^{(k)}$ but with every tensor product $V_i\otimes|i\rangle\langle i|
=(R^{k_1(i)}L^{k_2(i)}\cdots R^{k_{l-1}(i)}L^{k_{l}(i)})\rho_0(R^{k_1(i)}L^{k_2(i)}\cdots R^{k_{l-1}(i)}L^{k_{l}(i)})^*\otimes |i\rangle\langle i|$ removed whenever the product appearing in $V_i$ describes a sequence in which the walk has returned to zero at some moment. Alternatively, the paths returning to zero are removed at every iteration (via a projection to its complement, $I-|0\rangle\langle 0|$). For instance, if $\alpha$ is the state given by (\ref{secondst}), then
\begin{equation}
P\alpha=L^2\rho_0L^{2*}\otimes|-2\rangle\langle -2|+R^2\rho_0 R^{2*}\otimes|2\rangle\langle 2|\;.
\end{equation}
Now for a given state $\rho$, let $S_n(\rho)$ denote the probability of occurrence of a site other than zero at time $n$, that is,
\begin{equation}
S_n(\rho):=\mu(P\Omega^n(\rho)),
\end{equation}
where
\begin{equation}
\mu(\rho)=\mu(\sum_i \rho_i\otimes |i\rangle\langle i|):=\sum_i tr(\rho_i)
\end{equation}
Define the {\bf return probability} by
\begin{equation}\label{retprob}
R:=1-\lim_{n\to\infty} S_n(\rho)
\end{equation}
We say that site $|0\rangle$ is {\bf recurrent} for $\Omega$ if $R=1$ for all $\rho=\rho_0\otimes |0\rangle\langle 0|$, and we say it is {\bf transient} otherwise. We give the analogous definition for walks starting at sites other than $|0\rangle$.

\begin{remark}
When considering a quantum system together with the probabilistic notion of recurrence, one has to address the problem of how to describe a first return to the origin. In particular, if one has to check, at certain moments, whether a system has reached a certain state then such inspection is a measurement which modifies the system. Our approach here is the same as the one taken by Gr\"unbaum et al. \cite{werner}, that is, we consider a notion of recurrence that {\bf includes} the system monitoring into the description. This is not the only known possibility, as other definitions of recurrence can be found in the literature, see, e.g., \v Stefa\v n\'ak et al. \cite{jex}.
\end{remark}

\begin{remark}\label{types_recurrence}
We note that the definition above resembles a kind of {\bf site} recurrence, that is, recurrence of a particular position on $\mathbb{Z}$. One should compare this notion with {\bf state} recurrence, presented in \cite{bourg}.
\end{remark}

\begin{remark}
The definition of recurrence we presented makes sense for any OQRW, but in this section we will give emphasis on random walks induced by PQ-matrices. In the following section we will discuss a general definition of positive recurrence, for which PQ-induced OQRWs are an important class of examples.
\end{remark}

We recall some basic facts on classical random walks. Consider a random walk $\Omega$ on $\mathbb{Z}$, with $\Omega_0=0$. Let $T=\min\{n\geq 1: \Omega_n=0\}$ be the time of first return of the walk to its starting point. Then the number of paths of length $2k$ starting at 0 and first returning to 0 at time $t=2k$ is
\begin{equation}\label{bigcounting}
\alpha_{2k}=\frac{1}{2k-1}\binom{2k}{k},
\end{equation}
see \cite{grimmett}. In particular, for the symmetric random walk on $\mathbb{Z}$ ($p=q=1/2$), $P(T=2k)=\alpha_{2k} 2^{-2k}$. We would like to calculate $F(2N)$, the number of paths of length $2N$ starting at 0 and returning to 0 at some time between 1 and $2N$. The idea is simple: let $1< 2k\leq 2N$ (i.e., $k\in\{1,2,\dots, N\}$) and let $A_{2k}^{2N}$ denote the set of paths of length $2N$ starting at 0 and first returning to 0 at time $t=2k$. Note that $A_{2k}^{2N}\cap A_{2l}^{2N}=\emptyset$ if $k\neq l$. Then the number of paths of length $2N$ such that its first return to 0 occurs at $2k$ equals $\alpha_{2k}$ multiplied by the number of all possible paths of length $2N-2k$ starting at 0, which is $2^{2N-2k}$. Then we can conclude that $F(2N)=\sum_{k=1}^{N} \#A_{2k}^{2N}=\sum_{k=1}^{N} \alpha_{2k}2^{2N-2k}$.

\medskip

Now we note that the paths counted by $F(2N)$ (the ones returning to 0) are exactly those we discard when we compute the return probability via (\ref{retprob}). Thus, given $2N$ iterations, we are left with $\beta=2^{2N}-F(2N)$ paths. By (\ref{retprob}), if the sum of the probabilities of the $\beta$ paths mentioned goes to zero as $n\to\infty$, we conclude that $0$ is recurrent.

\medskip

{\bf Revisiting recurrence of the classical random walk on $\mathbb{Z}$.} We review this well-known probability result in our setting and notation, as some of the quantum cases we will consider are related to the reasoning made here. Also see \cite{attal}. Let $L=\sqrt{p}U_1,\;\;\; R=\sqrt{q}U_2$, where $p,q\in(0,1)$, $p+q=1$ and $U_i$, $i=1,2$ are order 2 unitary matrices. It is clear that $L$, $R$ satisfy (\ref{sum1}). We can show that, just like the classical case, the random walk induced by such matrices will be recurrent in the sense just described if and only if $p=q=1/2$. In fact, consider first the symmetric random walk. Then, by the above discussion, to establish recurrence one has to show that
\begin{equation}
\lim_{N\to\infty}[2^{2N}-F(2N)]\Big(\frac{1}{2}\Big)^{2N}=0
\end{equation}
Rewriting, we have
\begin{equation}
[2^{2N}-F(2N)]\Big(\frac{1}{2}\Big)^{2N}=1-\sum_{k=1}^N\frac{1}{2k-1}\binom{2k}{k}2^{-2k},
\end{equation}
so we are done, since
\begin{equation}
\sum_{k=1}^\infty\frac{1}{2k-1}\binom{2k}{k}2^{-2k}=1
\end{equation}
For the general case, we consider the following. Let $P_R^{2k}=\alpha_{2k}(pq)^k$ be the probability of first return to the origin in exactly $2k$ steps and let $P_{2N}$ be the probability of return in up to $2N$ steps. Then $P_{2N}=\sum_{k\leq N} P_R^{2k}$ so we get, by writing $\varphi(p)=p(1-p)$,
\begin{equation}\label{cl_calculation}
P_{2N}=\sum_{k\leq N} \frac{1}{2k-1}\binom{2k}{k}p^k(1-p)^k\leq \sum_{k=1}^\infty \frac{1}{2k-1}\binom{2k}{k}\varphi(p)^k\leq \sum_{k=1}^\infty \frac{1}{2k-1}\binom{2k}{k}\Big(\frac{1}{4}\Big)^k=1,
\end{equation}
and in the last inequality we have equality if and only if $p=1/2$.

\qee

We are interested in the following problem: given two matrices $L$, $R$ satisfying (\ref{sum1}), what conditions should be satisfied by its entries in order to obtain recurrence of the associated open quantum random walk? We will consider 1-qubit walk on $\mathbb{Z}$, with $L$ and $R$ diagonal or antidiagonal matrices describing one-step transitions to the left, or to the right, respectively.

\begin{remark}
We emphasize that, unlike the situation for the matrix representation of a given channel, the transition probabilities for an OQRW induced by Kraus matrices depend on the particular choice of such matrices. Because of this, to say that a walk is induced by a channel is ambiguous. So when we say that an OQRW is {\bf induced} by a PQ-channel $\Phi$, we mean that the channel is expressed in the form $\Phi(\rho)=\sum_i V_i\rho V_i^*$, where the $V_i$ are diagonal or antidiagonal (or, more generally, that the $V_i$ belong to $PQ_d$, see section \ref{secpq}) and then these particular matrices are taken as transitions.
\end{remark}

{\bf Example: Amplitude Damping.} As before, consider the initial state $\rho^{(0)}=\rho_0\otimes |0\rangle\langle 0|$. It is a simple matter to show that for the OQRW induced by the amplitude damping channel, the origin is transient. Indeed, let $L=V_1$ and $R=V_2$ be the Kraus matrices for such channel given in the appendix. First we note that
\begin{equation}
tr(L^n\rho_0 L^{n*})=\rho_{11}+(1-p)^n\rho_{22}
\end{equation}
and
\begin{equation}
tr(R^n\rho_0 R^{n*})=\left\{
\begin{array}{rl}
\rho_{22} p & \text{if } n=1,\\
0 & \text{if } n\geq 2\,.
\end{array} \right.\end{equation}
Also, it is clear that $V_1V_2\cdots V_n\rho_0 V_n^*V_{n-1}^*\cdots V_2^*V_1^*=0$ whenever at least two of the $V_i$ are equal to $R$, and we have
\begin{equation}
tr(L^rRL^s\rho_0L^{s*}R^*L^{r*})=p (1-p)^s\rho_{22},\;\;\; r\geq 0, s\geq 2
\end{equation}
(note that $s=0$ and $s=1$ correspond to trajectories that reach the zero position).
Thus
\begin{equation}
S_n(\rho_0)=tr(L^n\rho_0 L^{n*})+\sum_{k=2}^n tr(L^{n-k-1}RL^k\rho_0L^{k*}R^*L^{n-k-1*})=
\rho_{11}+(1-p)^n\rho_{22}+\sum_{k=2}^np (1-p)^k\rho_{22}
\end{equation}
which implies transience, since $\lim_{n\to\infty} S_n(\rho)=\rho_{11}+\rho_{22}(1-p)^2>0$.

\qee

The following lemma, the proof of which is a simple calculation, is the basic fact to be noticed and which will be used in the subsequent analysis.
\begin{lemma}\label{combinalemma} (Tracial behavior of OQRW induced by PQ-channels.)
Let $\rho=(\rho_{ij})$ density matrix. Then for all PQ-matrix $V$, $tr(V\rho V^*)$ does not depend on nondiagonal entries of $\rho$.
\end{lemma}
Now we consider 3 cases which illustrates the problem of recurrence of OQRW induced by 1-qubit unital PQ-channels with 2 Kraus matrices. Though the analysis below is a typical example of the combinatorial approach one may take on calculating probabilities in our setting, they are not essential for understanding the subsequent results, so this part might be skipped by the reader. The conclusion is contained in Theorem \ref{pro51} below.

\medskip

{\bf Case 1.} Consider
\begin{equation}\label{type1}
L=\begin{bmatrix} l_{11} & 0 \\ 0 & l_{22}\end{bmatrix},\;\;\; R=\begin{bmatrix} r_{11} & 0 \\ 0 & r_{22}\end{bmatrix}
\end{equation}
We suppose, as before, an initial state $\rho^{(0)}=\rho_0\otimes |0\rangle\langle 0|$, $\rho_0=(\rho_{ij})\in M_2(\mathbb{C})$ density matrix. For the first iteration, the probability of presence in $|-1\rangle$ is $tr(L\rho_0 L^*)=|l_{11}|^2\rho_{11}+|l_{22}|^2\rho_{22}$,
and the probability of presence in $|1\rangle$ is $tr(R\rho_0 R^*)=|r_{11}|^2\rho_{11}+|r_{22}|^2\rho_{22}$. Now we note that for the $n$-step transition probabilities, only the quantities of each direction (left or right) matters. In general, we have
\begin{equation}
tr(L^kR^l\rho_0 R^{l*}L^{k*})=(|l_{11}|^2)^k(|r_{11}|^2)^l\rho_{11}+(|l_{22}|^2)^k(|r_{22}|^2)^l\rho_{22}
\end{equation}
and the same is true for any sequence of $k$ matrices $L$ and $l$ matrices $R$. This provides some resemblance to the classical case. Let $P_R^{2k}=\alpha_{2k}tr(L^kR^k\rho_0R^{k*}L^{k*})$ be the probability of first return to the origin in $2k$ steps and let $P_{2N}$ be the probability of return in up to $2N$ steps. Then $P_{2N}=\sum_{k\leq N} P_R^{2k}$ so we get, recalling def. (\ref{bigcounting}),
$$P_{2N}=\sum_{k\leq N} \frac{1}{2k-1}\binom{2k}{k}tr(L^kR^k\rho_0R^{k*}L^{k*}) \leq \sum_{k=1}^\infty \frac{1}{2k-1}\binom{2k}{k}tr(L^kR^k\rho_0R^{k*}L^{k*})=$$
$$
=\sum_{k=1}^\infty \frac{1}{2k-1}\binom{2k}{k}\Big((|l_{11}|^2)^k(|r_{11}|^2)^k\rho_{11}+(|l_{22}|^2)^k(|r_{22}|^2)^k\rho_{22}\Big)=$$
\begin{equation}\label{q_calculation}
=\rho_{11}\sum_{k=1}^\infty\frac{1}{2k-1}\binom{2k}{k}(|l_{11}|^2)^k(|r_{11}|^2)^k+\rho_{22}\sum_{k=1}^\infty \frac{1}{2k-1}\binom{2k}{k}(|l_{22}|^2)^k(|r_{22}|^2)^k
\end{equation}
The summations above are of the same kind as the classical ones (see \eqref{cl_calculation}). So, noting that \eqref{sum1} implies $|r_{ii}|^2=1-|l_{ii}|^2$, $i=1,2$, we conclude that a $PQ$-channel of this case is recurrent if and only if $|l_{ii}|^2=|r_{ii}|^2=1/2$, $i=1,2$.
%Note that the assumption of unit preservation has not been used here.

\medskip

{\bf Case 2.} Consider
\begin{equation}\label{type2}
L=\begin{bmatrix} 0 & l_{12} \\ l_{21} & 0 \end{bmatrix},\;\;\; R=\begin{bmatrix} 0 & r_{12} \\ r_{21} & 0\end{bmatrix}
\end{equation}
Assuming that $L^*L+R^*R=I$, we see that this implies $|l_{12}|^2+|r_{12}|^2=1$ and $|l_{21}|^2+|r_{21}|^2=1$. We have
$$tr(L^{2k}R^{2l}\rho_0R^{2l*}L^{2k*})=(|l_{12}|^2)^k(|l_{21}|^2)^k(|r_{12}|^2)^k(|r_{21}|^2)^k\rho_{11}+(|l_{12}|^2)^k(|l_{21}|^2)^k(|r_{12}|^2)^k(|r_{21}|^2)^k\rho_{22}=$$
\begin{equation}
=(|l_{12}|^2)^k(|l_{21}|^2)^k(|r_{12}|^2)^k(|r_{21}|^2)^k
\end{equation}
The trace associated to a certain sequence is, as before, a number of the form $c_1\rho_{11}+c_2\rho_{22}$. With respect to $c_1$ we have the following: Let $V_{2N}1V_{2N-1}\cdots V_{1}$ be a sequence of $L$ and $R$'s. Then, $V_1=L$ will contribute with a coefficient $|l_{21}|^2$ and $V_1=R$ will contribute with a coefficient $|r_{21}|^2$. On the other hand, $V_2=L$ will contribute with a coefficient $|l_{12}|^2$ and $V_2=R$ will contribute with a coefficient $|r_{12}|^2$ and so on. That is, subscripts $21$ occur in odd instants of time, and subscripts $12$ occur in even instants of time. With respect to $c_2$ we have the analogous reasoning except that $12$ and $21$ are exchanged. Thus, for instance, $V_1=L$ and $V_2=R$ gives
\begin{equation}
tr(V_2V_1\rho_0V_1^*V_2^*)=tr(RL\rho L^*R^*)=c_1\rho_{11}+c_2\rho_{22}=|l_{21}|^2|r_{12}|^2\rho_{11}+|l_{12}|^2|r_{21}|^2\rho_{22}
\end{equation}
That is, we aggregate coefficients in the order $21, 12, 21, \dots$ for $\rho_{11}$ and  $12, 21, 12, \dots$ for $\rho_{22}$.
For $b=0,\dots, k$, let $p_{b,2k}$ denote the probability of occurrence of a path that first returns to the origin in $2k$ steps and does that by moving to the right in exactly b instants of time which are odd integers. Let $P_{b,2k}$ denote the set of all such paths. For instance, $LLRLLRRR$
is a sequence associated to a path in $P_{2,8}$, that is, $k=4$ and the path moves right in 2 instants which are odd integers (instants 3 and 7), besides moving right in instants 6 and 8, which we are not counting. Consider one such path in which $n_1=b$ moves to the right occurred at moments given by odd integers, $n_2$ moves to the right occurred at even moments, $n_3$ moves to the left occurred at odd moments, and $n_4$ occurred at even moments. If the path has length $2k$ then $\sum_i n_i=2k$. Also if the path first returns to the origin on $2k$ steps then we must have $k$ moves to the right and $k$ moves to the left, thus $n_1+n_2=n_3+n_4=k$ and $n_2=k-n_1$, $n_4=k-n_3$. Moreover since the path went right on exactly $n_1=b$ instants of time which are odd integers then the path went left on exactly $k-b$ instants of time which are odd integers. Hence this gives
$$p_{b,2k}=c_1\rho_{11}+c_2\rho_{22}=$$
\begin{equation}\label{probpb}
=(|r_{21}|^2)^b(|r_{12}|^2)^{k-b}(|l_{21}|^2)^{k-b}(|l_{12}|^2)^b\rho_{11}+(|r_{12}|^2)^b(|r_{21}|^2)^{k-b}(|l_{12}|^2)^{k-b}(|l_{21}|^2)^b\rho_{22}
\end{equation}
Note that $\sum_{b=0}^{k} \#P_{b,2k}=\alpha_{2k}$ and also that $\#P_{b,2k}=\#P_{k-b,2k}$. It is worth noting that some of the summands are equal to zero, e.g. $\#P_{0,2k}=\#P_{k,2k}$). For instance, let $k=4$ then the above equation gives
\begin{equation}
p_{1,8}=(|r_{21}|^2)(|r_{12}|^2)^{3}(|l_{21}|^2)^{3}(|l_{12}|^2)\rho_{11}+(|r_{12}|^2)(|r_{21}|^2)^{3}(|l_{12}|^2)^{3}(|l_{21}|^2)\rho_{22}
\end{equation}
\begin{equation}
p_{2,8}=(|r_{21}|^2)^2(|r_{12}|^2)^{2}(|l_{21}|^2)^{2}(|l_{12}|^2)^2\rho_{11}+(|r_{12}|^2)^2(|r_{21}|^2)^{2}(|l_{12}|^2)^{2}(|l_{21}|^2)^2\rho_{22}
\end{equation}
\begin{equation}
p_{3,8}=(|r_{21}|^2)^3(|r_{12}|^2)(|l_{21}|^2)(|l_{12}|^2)^3\rho_{11}+(|r_{12}|^2)^3(|r_{21}|^2)(|l_{12}|^2)(|l_{21}|^2)^3\rho_{22}
\end{equation}
Recalling that $|l_{21}|^2+|r_{21}|^2=1$ and $|l_{12}|^2+|r_{12}|^2=1$, write $x=|l_{21}|^2$ so $|r_{21}|^2=1-x$, and let $y=|r_{12}|^2$ so $|l_{12}|^2=1-y$, so we get
\begin{equation}
p_{1,8}=(x^3(1-x)y^3(1-y))\rho_{11}+(x(1-x)^3y(1-y)^3)\rho_{22}
\end{equation}
\begin{equation}
p_{2,8}=(x^2(1-x)^2y^2(1-y)^2)\rho_{11}+(x^2(1-x)^2y^2(1-y)^2)\rho_{22}
\end{equation}
\begin{equation}
p_{3,8}=(x(1-x)^3y(1-y)^3)\rho_{11}+(x^3(1-x)y^3(1-y))\rho_{22}
\end{equation}

Let $P_{2N}$ be the probability of return in up to $2N$ steps. Then
\begin{equation}\label{qsumm}
P_{2N}=\sum_{k\leq N}\sum_{b=0}^{k}\#P_{b,2k}p_{b,2k}
\end{equation}
We are interested in the limit of the above expression as $N$ goes to infinity. We note, once again, that if the entries of $L$, $R$ induce a classical symmetric distribution, i.e.,
\begin{equation}\label{csdist}
|l_{12}|^2=|l_{21}|^2=|r_{12}|^2=|r_{21}|^2=\frac{1}{2}
\end{equation}
then $p_{b,2k}=(1/4)^k$ for all $b$ and $k$, and then (\ref{qsumm}) produces the classical sum, so we have recurrence in this case. We claim that this is the only possibility. To see this, note that the numbers $\#P_{b,2k}$ are symmetric in $b$, so we just need to study the behavior of
\begin{equation}
f_k(x,y)=\sum_{b=0}^k \#P_{b,2k}x^b(1-x)^{k-b}y^b(1-y)^{k-b}
\end{equation}
Recalling that $\#P_{b,2k}=\#P_{k-b,2k}$, it is not difficult to show that the maximum for $f_k$ in $[0,1]\times[0,1]$ occurs only at $(1/2,1/2)$, so we are done. So if $\lim_{N\to\infty} P_{R,N}=1$ then the associated OQRW is recurrent and this occurs only for entries inducing the classical symmetric distribution (\ref{csdist}), else it is transient.

\medskip

{\bf Case 3.} Consider
\begin{equation}\label{type3}
L=\begin{bmatrix} l_{11} & 0 \\ 0 & l_{22}\end{bmatrix},\;\;\; R=\begin{bmatrix} 0 & r_{12} \\ r_{21} & 0\end{bmatrix}
\end{equation}
From $L^*L+R^*R=I$ we get $|l_{11}|^2+|r_{21}|^2=1$ and $|l_{22}|^2+|r_{12}|^2=1$. This case can be seen as a combination of the previous two. More precisely, whenever $R$ (the antidiagonal matrix) occurs, the contributions $r_{12}$ and $r_{21}$ alternate between $\rho_{11}$ and $\rho_{22}$ just as in case 2. On the other hand, whenever $L$ appears, it contributes a factor $|l_{11}|^2$ to $\rho_{11}$ and $|l_{22}|^2$ to $\rho_{22}$ unless the walk has previously moved to the right an odd number of times which, in such case,  $|l_{11}|^2$ contributes to $\rho_{22}$ and $|l_{22}|^2$ to $\rho_{11}$. For instance,
\begin{equation}
tr(RL^2RL\rho_0L^*R^*L^{2*}R^*)=|r_{12}|^2(|l_{22}|^2)^2|r_{21}|^2|l_{11}|^2\rho_{11}+|r_{21}|^2(|l_{11}|^2)^2|r_{12}|^2|l_{22}|^2\rho_{22}
\end{equation}
More generally, let $q_{c,2k}$ denote the probability of occurrence of a path that first returns to the origin in $2k$ steps, with $c$ being the number of times in which a move to the left has occurred after moving right an even amount of times. Let $Q_{c,2k}$ the set of all such paths. For instance, $LLRLRLRR$ is a sequence associated to a path in $Q_{3,8}$, that is, $k=4$ and the path moves left in 3 instants (times 1, 2 and 6) in which moving right has previously occurred an even amount of times (0, 0 and 2, respectively). Note that $\sum_{c=0}^k \#Q_{c,2k}=\alpha_{2k}$. Then $q_{c,2k}$ can be written as
\begin{equation}
q_{c,2k}=\Big[(|l_{11}|^2)^{n_1}(|l_{22}|^2)^{n_2}(|r_{12}|^2)^{n_3}(|r_{21}|^2)^{n_4})\Big]\rho_{11}+\Big[(|l_{11}|^2)^{n_5}(|l_{22}|^2)^{n_6}(|r_{12}|^2)^{n_7}(|r_{21}|^2)^{n_8})\Big]\rho_{22}
\end{equation}
Above $n_1=c$, the amount of times in which moving left occurred after moving right an even amount of times and $n_2$ is the amount of times in which moving left occurred after moving right an odd amount of times. Hence for a path first returning to zero in $2k$ steps we conclude that $n_1+n_2=k$ and so $n_2=k-n_1$. Similarly, $n_5=n_2$, $n_6=n_1$, $n_3+n_4=n_7+n_8=k$ and $n_3+n_7=k$. Hence,
\begin{equation}
q_{c,2k}=\Big[(|l_{11}|^2)^{c}(|l_{22}|^2)^{k-c}(|r_{12}|^2)^{k-n_4}(|r_{21}|^2)^{n_4}\Big]\rho_{11}+\Big[(|l_{11}|^2)^{k-c}(|l_{22}|^2)^{c}(|r_{12}|^2)^{n_7}(|r_{21}|^2)^{k-n_7}\Big]\rho_{22}
\end{equation}
Finally, note that if $k$ is even then $n_4$ equals $k/2$, otherwise $n_4=\lceil k/2 \rceil$, so
\begin{equation}\label{qc2kexp}
q_{c,2k}=(|l_{11}|^2)^{c}(|l_{22}|^2)^{k-c}(|r_{12}|^2)^{k-\lceil k/2 \rceil}(|r_{21}|^2)^{\lceil k/2 \rceil}\rho_{11}+(|l_{11}|^2)^{k-c}(|l_{22}|^2)^{c}(|r_{12}|^2)^{\lceil k/2 \rceil}(|r_{21}|^2)^{k-\lceil k/2 \rceil}\rho_{22}
\end{equation}
Write $x=|l_{11}|^2$ so $1-x=|r_{21}|^2$ and let $y=|l_{22}|^2$ so $1-y=|r_{12}|^2$. If the channel induced by the $V_i$ is unital then we must have $LL^*+RR^*=I$ which implies  $|l_{11}|^2+|r_{12}|^2=1$ and $|l_{22}|^2+|r_{21}|^2=1$ and so $x=y$. Then expression (\ref{qc2kexp}) becomes
\begin{equation}\label{qc2kexpsimp}
q_{c,2k}=x^k(1-x)^k\rho_{11}+x^k(1-x)^k\rho_{22}=x^k(1-x)^k
\end{equation}
Let $P_{2N}$ be the probability of return in up to $2N$ steps. Then under unitality assumption,
\begin{equation}\label{qsumm2}
P_{2N}=\sum_{k\leq N}\sum_{c=0}^{k}\#Q_{c,2k}q_{c,2k}=\sum_{k\leq N}x^k(1-x)^k\sum_{c=0}^{k}\#Q_{c,2k}=\sum_{k\leq N}\alpha_{2k}x^k(1-x)^k
\end{equation}
and so we conclude the walk is recurrent if and only if
\begin{equation}\label{gencriterion}
\lim_{N\to\infty} P_{2N}=1.
\end{equation}
As in cases 1 and 2, if the entries of $L$, $R$ induce the classical symmetric distribution (\ref{csdist}) then $q_{c,2k}=(1/4)^k$ for all $c$ and $k$, and then (\ref{qsumm2}) produces the classical sum and this is the only possibility since it is the only maximum of the function $x\mapsto [x(1-x)]^k$. Note that the nonunital case has to be analyzed separately, and is less straightforward than the one studied here. We have concluded:

\begin{theorem}\label{pro51}
(Recurrence of the nearest neighbor OQRW on $\mathbb{Z}$.) The OQRW on $\mathbb{Z}$ with initial state $\rho_0\otimes |0\rangle\langle 0|$ induced by a 1-qubit unital PQ-channel with 2 Kraus matrices $V_i$, $i=1,2$ is recurrent if and only if the nonzero entries of the $V_i$ have square moduli equal to $1/2$.
\end{theorem}

\begin{remark}
One might be interested in the nonunital case. Note that cases 1 and 2 are always unital. As for case 3, this remains an open question since expression (\ref{qc2kexpsimp}) is true only under the unitality assumption. In principle, one has to find the coefficients $\#Q_{c,2k}$ for every $c$ and $k$.
\end{remark}

\section{Stationarity and positive recurrence}\label{section_station}

In the previous section we have studied OQRWs and seen that we are able to define and examine recurrence in some concrete cases. Locally, such walks present a quantum (noncommutative) character, but its dissipative (open) structure suggests that in the long term this kind of system might reveal some behavior which is in some sense classical, not unlike the asymptotics of many quantum channels. In this section we will describe a notion of {\it positive} recurrence, in analogy to the definition seen in standard probability theory \cite{norris}. We are encouraged to proceed in this direction, partly due to the progress we made with respect to recurrence, and also because of the somewhat dual behavior open walks present. In this way we will examine a class of stochastic processes on which it is possible to study positive recurrence and which is strictly larger than classical probabilistic systems.

\begin{remark}\label{paratodos}
We remark that all results obtained in this section hold for general OQRWs, with open walks induced by PQ-matrices being just one class of examples.
\end{remark}

In classical probability we say that a measure $\mu$ is {\bf stationary} if
\begin{equation}\label{cl_stat}
\sum_x \mu(x)p(x,y)=\mu(y),
\end{equation}
for some transition matrix $p$. If $\mu$ is a probability, we say that $\mu$ is a stationary distribution. We are interested in quantum counterparts of this definition. More precisely, we consider a Hilbert space $\mathcal{H}\otimes\mathcal{K}$ where $\mathcal{H}$ is the space describing the degrees of freedom of a particle (e.g. $\mathcal{H}=\mathbb{C}^2$) standing at some site described by $\mathcal{K}$, e.g., a separable infinite dimensional space so each vector on a basis describes a site on $\mathbb{Z}$. Consider a given initial state on $\mathcal{H}\otimes\mathcal{K}$, $\rho=\sum_i \rho_i\otimes |i\rangle\langle i|$. By considering expression (\ref{cl_stat}), we want to establish an analogy between $\mu(x)$ and the probability associated to a site, e.g. $tr(\rho_i)$. Motivated by this we have the following definition.

\medskip

{\bf Definition.} If $\rho$ is a positive operator $\rho=\sum_i \rho_i\otimes |i\rangle\langle i|$, we say that $\rho$ is {\bf stationary}, or {\bf invariant}, with respect to the OQRW $\Omega$ if
\begin{equation}
\sum_j B_j^i\rho_j B_j^{i*}=\rho_i,\;\;\;\forall\; i\in\mathbb{Z}
\end{equation}
Multiply the above equation by $B_i^{i_1}$ and $B_i^{i_1*}$ on the left and right, respectively, and summing over $i$ we get
\begin{equation}
\sum_{i,j} B_i^{i_1}B_j^i\rho_j B_j^{i*}B_i^{i_1*}=\sum_i B_i^{i_1}\rho_iB_i^{i_1*}=\rho_{i_1}, \;\forall\; i\in\mathbb{Z}
\end{equation}
By induction, we can establish that if $\rho$ is stationary then
\begin{equation}\label{general_stationar}
\sum_{i_1,i_2,\dots, i_n} B_{i_n}^iB_{i_{n-1}}^{i_{n}}\cdots B_{i_1}^{i_2}\rho_{i_1}B_{i_1}^{i_2*}\cdots B_{i_{n-1}}^{i_{n}*}B_{i_n}^{i*}=\rho_i,\;\forall\; i\in\mathbb{Z}.
\end{equation}
We will be mostly interested in the case $\rho$ is a density matrix, but below we will need to reason on more general situations as well.

\medskip

We say that site $|j\rangle$ is {\bf accessible} from $|i\rangle$, $i\to j$, if for every nonzero $\rho_i$ there exists $(i,i_1,\dots, i_{n-1},j)$ such that
\begin{equation}
tr(B_{i_{n-1}}^{j}\cdots B_i^{i_1}\rho_i B_i^{i_1*}\cdots B_{i_{n-1}}^{j*})>0.
\end{equation}
In words, this means that the probability of reaching $j$ in finite time, starting from $i$, is positive. We say that $|i\rangle$ {\bf communicates} with $|j\rangle$ if $i\to j$ and $j\to i$. Notation: $i\leftrightarrow j$. It is an equivalence relation, just as in the classical case. If the set of all possible sites, e.g., $\{|i\rangle: i\in\mathbb{Z}\}$, consists of only one equivalence class, we say that the associated OQRW is {\bf irreducible}.

\medskip

The following Proposition illustrates a class property which is somewhat similar to the classical case. This fact will not be used in the subsequent proofs of this work, but we present it due to its independent interest.

\begin{pro}\label{prop_classe}
Suppose $|x\rangle$ is recurrent and $x\to y$. Then the probability of reaching $|x\rangle$, starting from $|y\rangle$, equals 1 and $|y\rangle$ is recurrent. In other words, recurrence is a class property.
\end{pro}
{\bf Proof.} If reaching $|x\rangle$ from $|y\rangle$ does not occur with probability 1 then the set of paths starting at $|x\rangle$, reaching $|y\rangle$ and then not returning to $|x\rangle$ in finite time has a strictly positive probability of occurring, which is impossible, since $|x\rangle$ is by assumption recurrent. Hence, reaching $|x\rangle$ from $|y\rangle$ occurs with probability 1. Now in order to show that $|y\rangle$ is recurrent, we consider a initial state $\rho=\rho_y\otimes |y\rangle\langle y|$ and perform the procedure given by definition, that is, apply $\Omega^n$ and then project on the complement of $|y\rangle$, that is, $I-|y\rangle\langle y|$. Now we observe the following: paths starting at $|y\rangle$ contain some $i_k=x$ with probability 1, because reaching $|x\rangle$ from $|y\rangle$ occurs with probability 1. Because of this, to project out a path which has reached $|y\rangle$ at some point removes, with probability 1, a path which has reached $|x\rangle$ at some point. Note that we can always write
\begin{equation}
B_{i_{n-1}}^{j}\cdots B_{i_{k-1}}^{x} \cdots B_y^{i_1}\rho_y B_y^{i_1*}\cdots B_{i_{k-1}}^{x}\cdots B_{i_{n-1}}^{j*}=B_{i_{n-1}}^{j}\cdots [B_{i_{k-1}}^{x} \cdots B_y^{i_1}\rho_y B_y^{i_1*}\cdots B_{i_{k-1}}^{x}]\cdots B_{i_{n-1}}^{j*}
\end{equation}
so the term on brackets can be seen as a initial density matrix at $|x\rangle$. Since $|x\rangle$ is by assumption recurrent, we conclude that the terms which have been removed (starting and returning to $|y\rangle$) carry the entire mass. So, as $\rho_y$ is arbitrary, we conclude that $|y\rangle$ is recurrent.

\qed

\begin{remark}
In the above theorem, the assumption that $x\to y$ is to be understood as implying that whenever we consider a local density $\rho_y\otimes|y\rangle$, then $\rho_y$ is a state which has been reached from $\rho_x\otimes|x\rangle$. In other words, the state where the process begins (in this case, $\rho_x\otimes|x\rangle$) determines all possible states in the future, implying that one is not allowed to consider states which cannot be reached by the process beginning at $\rho_x$. This is an important aspect of the random walk we are studying, one that does not appear in classical walks because of the absence of the positive matrix degree of freedom associated to classical sites. With this observation in mind, we say that an OQRW is {\bf recurrent} if it is irreducible and it contains some recurrent state. In this case, and because of the above Proposition, all states are recurrent and we have a quantum analog of a recurrent Markov chain.
\end{remark}

In classical probability one is interested in whether a recurrent state $i$ is {\it positive recurrent}. Let us recall this notion. Define $T_i$ as being the time of first return to $i$ and let
\begin{equation}\label{care11}
\gamma_j^i=\sum_{n=0}^\infty P_i(X_n=j,T_i>n)
\end{equation}
be the expected time spent in $j$ between visits to $i$. Define the expected return time \cite{durrett,norris} by
\begin{equation}\label{care12}
E_i(T_i)=\sum_j \gamma_j^i=\sum_j \sum_{n=0}^\infty P_i(X_n=j,T_i>n)
\end{equation}
Then we say that a recurrent state $i$ is positive recurrent if $E_i(T_i)<\infty$. The importance of this notion is the well-known fact: if a given Markov chain is such that some state is positive recurrent we conclude that there exists a stationary distribution, and conversely. In the classical symmetric walk on $\mathbb{Z}$, it is well-known that $E_i(T_i)=\infty$, that is, the origin is {\it null-recurrent}. It is a simple matter to construct examples of Markov chains which produce positive recurrent states, thus providing stationary distributions.

\medskip

We have seen that a definition of first return in quantum terms is possible by making use of projective measurements, and this has been employed in section \ref{oqrw} by adapting the construction in \cite{werner} to the open system studied here. Motivated by such facts and, up to some extent by (\ref{care11}) and (\ref{care12}), we would like to describe a version of positive recurrence in the context of open quantum walks, so we make use of the following definitions.

\medskip

{\bf Definition.} For a fixed initial state $\rho_x\otimes |x\rangle\langle x|$, let $S_{\rho_x,j}^T$  be the operator given by
\begin{equation}
S_{\rho_x,j}^T:=\sum_{i_1,\dots, i_{T-1}\neq x}B_{i_{T-1}}^{j}\cdots B_x^{i_1}\rho_x B_x^{i_1*}\cdots B_{i_{T-1}}^{j*}
\end{equation}
where $T \geq 1$ and $j\in \mathbb{Z}$.
Note that $S_{\rho_x,j}^1 = B_x^j \rho_x B_x^{j*}$. Also define
\begin{equation}\label{estac_eh_o_lim_iter}
\rho_{st,\rho_x}:=\sum_j\rho_{st,\rho_x}(j)\otimes |j\rangle\langle j|,\;\;\;\rho_{st,\rho_x}(j):=\sum_{T=1}^\infty S_{\rho_x,j}^{T}
\end{equation}

{\bf Definition.}
We will say that $|x\rangle$ is a {\bf positive recurrent} site if it is recurrent and there exists $\rho_x$ such that \begin{equation}\label{tracofinito}
\sum_j tr(\rho_{st,\rho_x}(j))<\infty,
\end{equation}
and
\begin{equation}\label{cond_imp}
\sum_{T=1}^\infty S_{\rho_x,x}^T=\rho_x.
\end{equation}

It is instructive to note the similarities between the classical definition of positive recurrence (via equation (\ref{care12})) and the new definition given above (via (\ref{estac_eh_o_lim_iter}) and (\ref{tracofinito})).

\begin{remark}
Note that eq. (\ref{cond_imp}) is automatically true in the classical case (open walk induced by  matrices which are multiples of the identity), but we are able to construct nonclassical examples satisfying (\ref{cond_imp}) as well.
\end{remark}

\begin{remark}
It is worth remarking one aspect concerning our definitions of recurrence and positive recurrence. As described in section \ref{oqrw}, we have defined recurrence of a site which imposes a condition on {\bf all} local density matrices (equation (\ref{retprob})). On the other hand,  the definition of positive recurrence is conditioned to the existence of {\bf some} density (satisfying (\ref{tracofinito}) and (\ref{cond_imp})). One might be interested in variations of these definitions. For other quantum applications, one could instead define recurrence of a particular site with respect to a fixed local density matrix. Under these terms, we would have a definition which is perhaps closer to what is presented in \cite{werner}, where the authors discuss recurrence of a pair $(U,\phi)$, $U:\mathcal{H}\to\mathcal{H}$ a unitary evolution on a Hilbert space $\mathcal{H}$, $\phi\in \mathcal{H}$ some given state.
\end{remark}

\begin{pro}\label{theo_ineq}
Let $\Omega$ be a OQRW and let $\lambda=\sum_j\lambda_j\otimes |j\rangle\langle j|$ be a stationary operator.
Then for any $j,k\in \mathbb{Z}$ we have the operator inequality $\rho_{st,\lambda_k}(j)\leq \lambda(j)$.
\end{pro}

{\bf Proof.} Fix $k \in \mathbb{Z}$. For every $j$ we have
\begin{equation}
\lambda_j=\sum_{i_0} B_{i_0}^j\lambda_{i_0}B_{i_0}^{j*}=\sum_{i_0\neq k}B_{i_0}^j\lambda_{i_0}B_{i_0}^{j*}+B_{k}^j\lambda_kB_{k}^{j*}=
\end{equation}
(repeat this procedure for $\lambda_{i_0}$ and so on)
$$=\sum_{i_0,i_1\neq k} B_{i_0}^jB_{i_1}^{i_0}\lambda_{i_1}B_{i_1}^{i_0*}B_{i_0}^{j*}+\Big(\sum_{i_0\neq k}B_{i_0}^jB_k^{i_0}\lambda_kB_k^{i_0*}B_{i_0}^{j*}  +B_{k}^j\lambda_kB_{k}^{j*}\Big)=\cdots =$$
\begin{equation}
\cdots=\sum_{i_0,i_1,\dots, i_n\neq k}B_{i_0}^jB_{i_1}^{i_0}\cdots B_{i_{n}}^{i_{n-1}}\lambda_{i_n}B_{i_{n}}^{i_{n-1}*}\cdots B_{i_1}^{i_0*}B_{i_0}^{j*}\;+
\end{equation}
$$+\;\Big(\sum_{i_0,\dots,i_{n-1}\neq k} B_{i_0}^jB_{i_1}^{i_0}\cdots B_{k}^{i_{n-1}}\lambda_kB_{k}^{i_{n-1}*}\cdots B_{i_1}^{i_0*}B_{i_0}^{j*} +\cdots+\sum_{i_0\neq k}B_{i_0}^jB_k^{i_0}\lambda_kB_k^{i_0*}B_{i_0}^{j*}+B_{k}^j\lambda_kB_{k}^{j*}\Big)\geq$$
$$\geq \sum_{i_0,\dots,i_{n-1}\neq k} B_{i_0}^jB_{i_1}^{i_0}\cdots B_{k}^{i_{n-1}}\lambda_kB_{k}^{i_{n-1}*}\cdots B_{i_1}^{i_0*}B_{i_0}^{j*} +\cdots+\sum_{i_0\neq k}B_{i_0}^jB_k^{i_0}\lambda_kB_k^{i_0*}B_{i_0}^{j*}+B_{k}^j\lambda_kB_{k}^{j*}=\sum_{T=1}^n S_{\lambda_k,j}^T.$$
As a consequence,
\begin{equation}\label{513}
\lim_{n\to\infty}\sum_{T=1}^n S_{\lambda_k,j}^T=\sum_{T=1}^\infty S_{\lambda_k,j}^T=:\rho_{st,\lambda_k}(j)\leq \lambda_j.
\end{equation}

\qed

\begin{theorem}\label{durrett1}
Suppose $|x\rangle$ is a positive recurrent site for a given OQRW. Then $\rho_{st,\rho_x}$ defined in \eqref{estac_eh_o_lim_iter}, satisfying (\ref{tracofinito}) and (\ref{cond_imp}), is a stationary operator.
\end{theorem}
{\bf Proof.} We must prove that
\begin{equation}
\sum_j B_j^i\rho_{st,\rho_x}(j)B_j^{i*}=\rho_{st,\rho_x}(i),\;\;\;\forall i\in\mathbb{Z}
\end{equation}
First suppose that $i = x$, then we have
$$\sum_j B_j^x\rho_{st,\rho_x}(j)B_j^{x*}=B_x^x\rho_{st,\rho_x}(x)B_x^{x*}+\sum_{j\neq x} \sum_{T=1}^\infty B_j^x S_{\rho_x,j}^{T}B_j^{x*}$$
\begin{equation}
= B_x^x\rho_x B_x^{x*}+\sum_{T=1}^\infty S_{\rho_x,x}^{T+1}=\rho_{st,\rho_x}(x).
\end{equation}
Now, if $i\neq x$, then
$$\sum_j B_j^i\rho_{st,\rho_x}(j)B_j^{i*}= B_x^i \sum_{T=1}^\infty S_{\rho_x,x}^T B_x^{i*}   + \sum_{j\neq x}\sum_{T=1}^\infty B_j^i S_{\rho_x,j}^{T}B_j^{i*}=B_x^i \sum_{T=1}^\infty  S_{\rho_x,x}^T B_x^{i*}   +\sum_{T=1}^\infty  S_{\rho_x,i}^{T+1}=$$
\begin{equation}
=B_x^i \rho_x B_x^{i*}   +\sum_{T=2}^\infty  S_{\rho_x,i}^{T}=\sum_{T=1}^\infty  S_{\rho_x,i}^{T}=\rho_{st,\rho_x}(i).
\end{equation}

\qed

Now we prove our main result on positive recurrence of irreducible OQRWs. The structure of the proof is inspired by the well-known result on Markov chains \cite{grimmett,norris}. As a consequence, and in a similar way as in the case of recurrence, we have that positive recurrence is a class property:

\begin{theorem}
Let $\Omega$ be an irreducible recurrent OQRW. The following are equivalent:

a) Every site is positive recurrent.

 b) Some site is positive recurrent.

 c) $\Omega$ has a stationary state $\pi=\sum_j \pi_j\otimes|j\rangle\langle j|$, $\sum_j tr(\pi_j)=1$.

\end{theorem}

{\bf Proof.} a) $\Rightarrow$ b) is immediate.

\medskip

b) $\Rightarrow $ c): Suppose $|i\rangle$ is positive recurrent. Then by Theorem \ref{durrett1}, $\rho_{st,\rho_i}$ is an invariant operator for some $\rho_i$. But if
\begin{equation}
\sum_j tr(\rho_{st,\rho_i}(j))=\sum_j\sum_{T=1}^\infty tr(S_{\rho_i,j}^T)<\infty
\end{equation}
then $\pi_k=\sum_{T=1}^\infty S_{\rho_i,k}^T/(\sum_j\sum_{T=1}^\infty tr(S_{\rho_i,j}^T))$ defines a stationary state (density matrix).

\medskip

c) $\Rightarrow$ a): let $|x\rangle$ be a site. Note that $\Omega$ is irreducible by assumption so stationarity of $\pi$ gives
\begin{equation}
\pi_x=\sum_{i_1,\dots i_{n-1}} B_{i_{n-1}}^x\cdots \pi_{i_1}\cdots B_{i_{n-1}}^{x*}>0.
\end{equation}
Let $\lambda_i:=\pi_i/tr(\pi_x)$. Then $\lambda_x:=\sum_i\lambda_i\otimes |i\rangle\langle i|$ is an invariant operator. By Proposition \ref{theo_ineq}, $\sum_{T=1}^\infty S_{\lambda_x,i}^T\leq \lambda_i$. Hence,
\begin{equation}
\sum_i tr(\rho_{st,\lambda_x}(i))=\sum_i tr(\sum_{T=1}^\infty S_{\lambda_x,i}^T)\leq\sum_i \frac{tr(\pi_i)}{tr(\pi_x)}=\frac{1}{tr(\pi_x)}<\infty
\end{equation}
Then note that $tr(\lambda_x)=tr(\sum_T S_{\lambda_x,x}^T)=1$, the last equality due to recurrence of $| x\rangle$. But then $tr(\lambda_x-\rho_{st,\lambda_x}(x))=0$ and the operator $\lambda_x-\rho_{st,\lambda_x}(x)$ is positive. Hence $\lambda_x=\rho_{st,\lambda_x}(x)$. This proves positive recurrence of $x$.

\qed

{\bf Example: Quantum random walks with retaining barrier.} We let $B_0^1=I$ and for $i\geq 1$, let
\begin{equation}
B_{i}^{i-1}=\begin{bmatrix} \sqrt{q_{11}} & 0 \\ 0 & \sqrt{q_{22}}\end{bmatrix} ,\;\;\; B_i^{i+1}=\begin{bmatrix} \sqrt{p_{11}} & 0 \\ 0 & \sqrt{p_{22}}\end{bmatrix},
\end{equation}
with $p_{jj},\; q_{jj}\geq 0$, $p_{jj}<q_{jj}$, $j=1, 2$ and assume that
$$B_i^{i-1*}B_i^{i-1}+B_i^{i+1*}B_i^{i+1}=\begin{bmatrix} p_{11}+q_{11} & 0 \\ 0 & p_{22}+q_{22}\end{bmatrix}=I.$$
This example can be seen as two copies of a random walk on $\mathbb{Z}$ with a retaining barrier at zero. Both walks move from $|0\rangle$ to $|1\rangle$ with probability 1 and in future moments it tends to move left with larger probability than moving right. Let $\rho=\rho_0\otimes |0\rangle\langle 0|$, $tr(\rho_0)=1$, then $B_0^{1}\rho_0 B_0^{1*}=\rho_0\otimes |1\rangle\langle 1|$ and for $i\geq 1$,
\begin{equation}
B_i^{i+1}\rho_0 B_i^{i+1*}=\begin{bmatrix}  p_{11}\rho_{11} & \sqrt{p_{11}p_{22}}\rho_{12} \\ \sqrt{p_{11}p_{22}}\ov{\rho_{12}} & p_{22}\rho_{22} \end{bmatrix},\;\;\; B_i^{i-1}\rho_0 B_i^{i-1*}=\begin{bmatrix} q_{11}\rho_{11} & \sqrt{q_{11}q_{22}}\rho_{12} \\ \sqrt{q_{11}q_{22}}\ov{\rho_{12}} & q_{22}\rho_{22} \end{bmatrix},
\end{equation}
\begin{equation}
B_{i+1}^iB_i^{i+1}\rho_0 B_i^{i+1*}B_{i+1}^{i*}=\begin{bmatrix}  p_{11}q_{11}\rho_{11} & \sqrt{p_{11}p_{22}}\sqrt{q_{11}q_{22}}\rho_{12} \\ \sqrt{q_{11}q_{22}}\sqrt{p_{11}p_{22}}\ov{\rho_{12}} & p_{22}q_{22}\rho_{22} \end{bmatrix}=B_{i-1}^iB_i^{i-1}\rho_0 B_i^{i-1*}B_{i-1}^{i*}
\end{equation}
In general,
$$B_{i_{n-1}}^{i_n}B_{i_{n-2}}^{i_{n-1}}\cdots B_{i_1}^{i_2}B_{0}^{i_1}\rho_0 B_{0}^{i_1*}B_{i_1}^{i_2*}\cdots B_{i_{n-2}}^{i_{n-1*}}B_{i_{n-1}}^{i_n*}=\begin{bmatrix}  p_{11}^{k-1}q_{11}^{n-k}\rho_{11} & \sqrt{p_{11}p_{22}}^{k-1}\sqrt{q_{11}q_{22}}^{n-k}\rho_{12} \\ \sqrt{q_{11}q_{22}}^{k-1}\sqrt{p_{11}p_{22}}^{n-k}\ov{\rho_{12}} & p_{22}^{k-1}q_{22}^{n-k}\rho_{22} \end{bmatrix},$$
where $k$ is the number of times the walk has moved right (note that above we write $p^{k-1}$, and not $p^k$, since the first move is to the right with probability one). Then by a classical argument $tr(\rho_{st,\rho_0}(0))=tr(\rho_0)=1$. We can also conclude that $\sum_i \rho_{st,\rho_x}(i)<\infty$, since both walks with barriers (from each position in the main diagonal) are known to be positive recurrent, for which a stationary distribution can be obtained \cite{grimmett}.

\qee

\begin{remark}\label{remark_ree}
It is clear that some aspects of the previous example are essentially the same as what is found in classical Markov chain examples, but written in a different notation. If the OQRW considered is one induced by PQ-matrices, then this resemblance is typical, i.e., conclusions may be obtained via the diagonal entries of a local density (i.e., a probability vector) and a calculation with entries of a classical stochastic matrix. However, this does not occur in general for OQRWs: if we consider a walk induced by non-PQ matrices, we see that questions on recurrence may depend on nondiagonal entries of local density matrices as well (recall Remark \ref{paratodos}). Also, this leads to an open question we briefly discuss in the next section.
\end{remark}

\section{Open questions}\label{openq}

A question on recurrence has been presented in \cite{attal}: can we obtain a recurrence criteria for an OQRW  based solely on the entries of the matrices defining the walk? Theorem \ref{pro51} shows that the answer is positive for walks induced by any unital 1-qubit PQ-channel with 2 Kraus matrices, and this shows some resemblance to the classical case. The question remains open for 1-qubit PQ-channels with 3 or 4 Kraus matrices (or even more matrices belonging to $PQ_2$, see section \ref{secpq}), and also for d-dimensional PQ-channels, $d>2$. Understanding the structure of such problems might be of assistance when considering the recurrence problem for OQRWs induced by general CPT maps.

\medskip

Closely related to our setting, one might consider a larger class of matrices, namely the ones on $PQ_d$ together with matrices in which an entire column is nonzero. For instance, take
\begin{equation}
V_1=\begin{bmatrix} v_{11} & 0 \\ 0 & v_{22} \end{bmatrix},\;\;\; V_2=\begin{bmatrix} 0 & v_{12}\\ v_{21} & 0 \end{bmatrix},\;\;\; V_3=\begin{bmatrix} w_{11} & 0 \\ w_{21} & 0\end{bmatrix},\;\;\; V_4=\begin{bmatrix} 0 & z_{12} \\ 0 & z_{22}\end{bmatrix}
\end{equation}
and suppose $\sum_i V_i^*V_i=I$. This induces a channel $\Phi$ which has a matrix representation of the form
\begin{equation}
[\Phi]=\begin{bmatrix} \times &  &  & \times \\ \times & \times & \times & \times \\ \times & \times & \times & \times \\ \times &  &  & \times \end{bmatrix},
\end{equation}
where the crosses denote nonzero numbers and the blank entries are zeroes. In this case coherences are generated by the main diagonal entries as well and so a density matrix may evolve into more complicated forms. Then one might ask the same questions on recurrence we made so far. Let $\alpha PQ_2$ denote the set of order 2 PQ-matrices together with matrices of the kind $V_3$ and $V_4$. Then we have that Lemma \ref{combinalemma} holds for the class $\alpha PQ_2$. One could also ask analogous questions about the class of matrices given by the ones on $PQ_2$  together with matrices in which an entire {\it row} is nonzero. Call $\beta PQ_2$ such class, which induces quantum channels such that classical entries of a density matrix (main diagonal) are induced by the coherences, i.e., the opposite phenomena of class $\alpha PQ_2$.

\medskip

As a related problem, one may consider the master equation associated to an open system \cite{ander,nielsen},
\begin{equation}
\frac{d\rho}{dt}=-\frac{i}{\hbar}[H,\rho]+\sum_j[2L_j\rho L_j^* -\{L_j^*L_j,\rho\}],
\end{equation}
Since solutions for such equations can be written in terms of completely positive maps (CP), we may look for a characterization of $H$ and the Lindblad operators $L_i$ so that a channel which is a solution can be described by permutations of diagonal matrices (i.e. a PQ-channel).

\medskip

Another question is the following. Recall the Remark \ref{remark_ree}, where we pointed out the resemblance of certain calculations of the two walks with a barrier with known calculations seen in classical probability. One could ask whether positive recurrence could be described simply by certain well-known classical processes. More precisely, can the properties of positive recurrence be directly inferred from the Markov chain induced by a stochastic matrix on the level of the diagonal entries of local density matrices? With respect to OQRWs induced by PQ-matrices, this is seen to be true, even if the local densities $\rho_x$ satisfying (\ref{tracofinito}) and (\ref{cond_imp}) are not diagonal. However, this does not occur for general OQRWs: consider again the two quantum walks with a barrier but such that in some site a rotation occurs, thus mixing diagonal entries with nondiagonal ones. It is an open question to understand how recurrence and positive recurrence are affected by such conditions. Concluding, we may ask: can one prove recurrence and positive recurrence for more general classes of channels, for instance, open quantum random walks where some (or all) matrices are not PQ-matrices? One simple example of OQRW on $\mathbb{Z}$ induced by non-PQ matrices, for which recurrence is easily proven, is given by the matrices
\begin{equation}
B_i^{i+1}=B=\frac{1}{\sqrt{2}}\begin{bmatrix} 1 & 1 \\ 0 & 0 \end{bmatrix},\;\;\;B_i^{i-1}=C=\frac{1}{\sqrt{2}}\begin{bmatrix} 0 & 0 \\1 & -1 \end{bmatrix},\;\;\;i\in\mathbb{Z}
\end{equation}
We note that $B+C$ equals the Hadamard matrix. This is an example of the class of OQRW induced by a unitary matrix $U=B+C$, with $C^*B=0$, for which a realization of a unitary quantum walk can be done \cite{attal}.

\section{Appendix: matrix representations for PQ-channels}

For convenience we write the matrix representations for some of the PQ-channels mentioned in this work. Channels 1-4 have been described in \cite{nielsen}, and for those we assume $p\in (0,1)$. The 2-qubit CNOT is described in \cite{novotny1}.

\medskip

1. Bit-flip, $\Phi_{bf}(\rho)=V_1\rho V_1^*+V_2\rho V_2^*$,
\begin{equation}
V_1=\sqrt{p}\begin{bmatrix} 1 & 0 \\ 0 & 1\end{bmatrix}, \;\;\; V_2=\sqrt{1-p}\begin{bmatrix} 0 & 1 \\ 1 & 0 \end{bmatrix},\;\;\; [\Phi_{bf}]=\begin{bmatrix} p & 0 & 0 & 1-p \\ 0 & p &  1-p & 0 \\ 0 & 1-p & p & 0 \\ 1-p & 0 & 0 & p \end{bmatrix}
\end{equation}

\medskip

2. Bit-phase-flip, $\Phi_{bpf}(\rho)=V_1\rho V_1^*+V_2\rho V_2^*$,
\begin{equation}
V_1=\sqrt{p}\begin{bmatrix} 1 & 0 \\ 0 & 1\end{bmatrix}, \;\;\; V_2=\sqrt{1-p}\begin{bmatrix} 0 & -i \\ i & 0 \end{bmatrix},\;\;\;[\Phi_{bpf}]=\begin{bmatrix} p & 0 & 0 & 1-p \\ 0 & p &  -1+p & 0 \\ 0 & -1+p & p & 0 \\ 1-p & 0 & 0 & p \end{bmatrix}
\end{equation}

\medskip

3. Amplitude damping, $\Phi_{ad}(\rho)=V_1\rho V_1^*+V_2\rho V_2^*$,
\begin{equation}
V_1=\begin{bmatrix} 1 & 0 \\ 0 & \sqrt{1-p}\end{bmatrix}, \;\;\; V_2=\begin{bmatrix} 0 & \sqrt{p} \\ 0 & 0 \end{bmatrix},\;\;\; [\Phi_{ad}]=\begin{bmatrix} 1 & 0 & 0 & p \\ 0 & \sqrt{1-p} &  0 & 0 \\ 0 & 0 & \sqrt{1-p} & 0 \\ 0 & 0 & 0 & 1-p \end{bmatrix}
\end{equation}

\medskip

4. Depolarizing channel, $\Phi_{dc}(\rho)=\sum_{i=1}^4 V_i\rho V_i^*$,
\begin{equation}
V_1=\sqrt{1-\frac{3p}{4}}\begin{bmatrix} 1 & 0 \\ 0 & 1 \end{bmatrix},\;\;\; V_2=\frac{\sqrt{p}}{2}\begin{bmatrix} 0 & 1\\ 1 & 0 \end{bmatrix},\;\;\; V_3=\frac{\sqrt{p}}{2}\begin{bmatrix} 0 & -i \\ i & 0\end{bmatrix},\;\;\; V_4=\frac{\sqrt{p}}{2}\begin{bmatrix} 1 & 0 \\ 0 & -1\end{bmatrix}
\end{equation}
\begin{equation}
[\Phi_{dc}]=\begin{bmatrix} 1-\frac{p}{2} & 0 & 0 & \frac{p}{2} \\ 0 & 1-p & 0 & 0 \\ 0 & 0 & 1-p & 0 \\ \frac{p}{2} & 0 & 0 & 1-\frac{p}{2}\end{bmatrix}
\end{equation}

\medskip

5. Two-qubit CNOT, $\Phi_{CN}(\rho)=C_1\rho C_1^*+C_2\rho C_2^*$, $p+r=1$
\begin{equation}
C_1=\sqrt{p}\begin{bmatrix} 1 & 0 & 0 & 0 \\ 0 & 1 & 0 & 0 \\ 0 & 0 & 0 & 1 \\ 0 & 0 & 1 & 0 \end{bmatrix}, \;\;\; C_2=\sqrt{r}\begin{bmatrix} 1 & 0 & 0 & 0 \\ 0 & 0 & 0 & 1 \\ 0 & 0 & 1 & 0 \\ 0 & 1 & 0 & 0 \end{bmatrix}
\end{equation}

\begin{equation}\label{largecnot}
[\Phi_{CN}]=\left(\begin{array}{cccccccccccccccc}
{\bf 1} &  &  &  &  & {\bf 0} &  &  &  &  & {\bf 0} &  &  &  &  & {\bf 0} \\    %1
 &  {p} &  {0} &  {r} &  {0} &  &  {0} &  {0} & {0} &  {0} &  &  {0} &  {0} &  {0} &  {0} &  \\  %2
 &  {0} &  {r} &  {p} &  {0} &  &  {0} &  {0} &  {0} &  {0} &  &  {0} &  {0} &  {0} &  {0} &  \\  %3
 &  {r} &  {p} &  {0} &  {0} &  &  {0} &  {0} &  {0} &  {0} &  &  {0} &  {0} &  {0} &  {0} &  \\  %4
 &  {0} &  {0} &  {0} &  {p} &  &  {0} &  {0} &  {0} &  {0} &  &  {0} &  {r} &  {0} &  {0} &  \\    %5
{\bf 0} &  &  &  &  & {\bf p} &  &   &   &   & {\bf 0} &   &   &   &   & {\bf r} \\    %6
 & 0 & 0 & 0 & 0 &  & 0 & p & 0 & 0 &  & 0 & 0 & 0 & r &  \\    %7
 & 0 & 0 & 0 & 0 &  & p & 0 & 0 & 0 &  & 0 & 0 & r & 0 &  \\    %8
 & 0 & 0 & 0 & 0 &  & 0 & 0 & r & 0 &  & 0 & p & 0 & 0 &  \\    %9
 & 0 & 0 & 0 & 0 &  & 0 & 0 & 0 & 0 &  & r & 0 & p & 0 &  \\    %10
{\bf 0} &   &   &   &   & {\bf 0} &   &   &   &   & {\bf r} &   &   &   &   & {\bf p} \\    %11
 & 0 & 0 & 0 & 0 &  & 0 & 0 & 0 & r &  & 0 & 0 & 0 & p &  \\    %12
 & 0 & 0 & 0 & r &  & 0 & 0 & p & 0 &  & 0 & 0 & 0 & 0 &  \\    %13
 & 0 & 0 & 0 & 0 &  & 0 & r & 0 & p &  & 0 & 0 & 0 & 0 &  \\    %14
 & 0 & 0 & 0 & 0 &  & r & 0 & 0 & 0 &  & p & 0 & 0 & 0 &  \\    %15
{\bf 0} &   &   &   &   & {\bf r} &   &   &   &   & {\bf p} &   &   &   &   & {\bf 0}       %16
\end{array}\right)
\end{equation}
The blank entries for $[\Phi_{CN}]$ are zeroes, which have been omitted for an easier visualization. The 2-qubit CNOT channel has been studied by Novotn\'y et al. \cite{novotny1} and illustrates, among other facts, the use of a convergence theorem for mixed-unitary channels proved by these authors.

\end{document}